\documentclass[fleqn,10pt]{wlscirep}
\usepackage[utf8]{inputenc}
\usepackage[T1]{fontenc}
\usepackage{amsfonts, multirow}
\usepackage[normalem]{ulem}
\usepackage{pdfpages}

 \colorlet{revisions}{orange!70!black}

\newcommand{\rev}[1]{{\color{black}#1}}
\newcommand{\revrev}[1]{{\color{black}#1}}

\title{Multilayer networks characterize human-mobility patterns by industry sector for the 2021 Texas winter storm}

\author[1,2,*]{Melissa Butler}
\author[1,2]{Alisha Khan}
\author[1,3,]{Francis Osei Tutu Afrifa}
\author[4,]{Yingjie Hu}
\author[1,2,*]{Dane Taylor}
\affil[1]{School of Computing, University of Wyoming, Laramie, WY, 82072, USA}
\affil[2]{Department of Mathematics \& Statistics, University of Wyoming, Laramie, WY, 82072, USA}
\affil[3]{Department of Atmospheric Science, University of Wyoming, Laramie, WY, 82072, USA}
\affil[4]{Department of Geography, University at Buffalo, Buffalo, NY, 14260, USA}

\affil[*]{Corresponding authors: mbutle15@uwyo.edu and dane.taylor@uwyo.edu}

\begin{abstract}

\rev{
Understanding human mobility during disastrous events is crucial for emergency planning and disaster management. We develop a methodology to construct time-varying, multilayer networks where edges encode observed movements between spatial regions (census tracts) and network layers encode movement categories by industry sectors (e.g., schools, hospitals). Using the 2021 Texas winter storm as a case study, we find that people markedly reduced movements to ambulatory healthcare services, restaurants, and schools, but prioritized movements to grocery stores and gas stations. Additionally, we study the predictability of nodes’ in- and out-degrees in the multilayer networks, which encode movements into and out of census tracts. Inward movements prove harder to predict than outward movements, especially during the storm. \revrev{Our} findings 
on the  reduction, prioritization, and predictability of sector-specific movements 
aim to  support mobility-related decisions during future extreme weather events. 
}
\end{abstract}

\begin{document}
\flushbottom
\maketitle

%---------------------------------------------------------------------------------%---------------------------------------------------------------------------------
\section{Introduction}
%---------------------------------------------------------------------------------
%---------------------------------------------------------------------------------

%
Networks encoding the spatio-temporal patterns of human movements (i.e., mobility networks) have  been developed and used to provide insights about daily commuting patterns \cite{gonzalez2008understanding,louail2015uncovering},    improve public transit infrastructures   \cite{louf2014congestion},   develop data-driven models for epidemic spreading \cite{meloni2011modeling,tizzoni2014use}, and   reveal geographic insights about segregation \cite{nilforoshan2023human} and  inequality \cite{xu2025using} (e.g., with respect to   access to goods and services). Of note, multilayer networks \cite{mucha2010community,kivela2014multilayer,bianconi2018multilayer} have been adopted as a leading framework for  mobility modeling, whereby different network layers have been utilized to represent different types of interconnected networks. Examples include networks that distinguish different modes of transportation \cite{de2014navigability,taylor2015topological,chodrow2016demand} or complementary  infrastructures within a single mode of transportation (e.g., different airlines \cite{cardillo2013emergence,taylor2021tunable}). Different layers can also be used to represent different  sources of data for mobility \cite{belyi2017global}, and it's worth noting that one might expect each mobility network layer to adhere to  different spatial and temporal constraints \cite{barthelemy2011spatial}. 
%
%\rev{We introduce a multilayer network structure that encodes movements across census tracts, prioritizing movements categorized by industry sector. } DRT: this is the next paragraph.

%

\begin{figure}[t!]
\centering
\includegraphics[width=\linewidth]{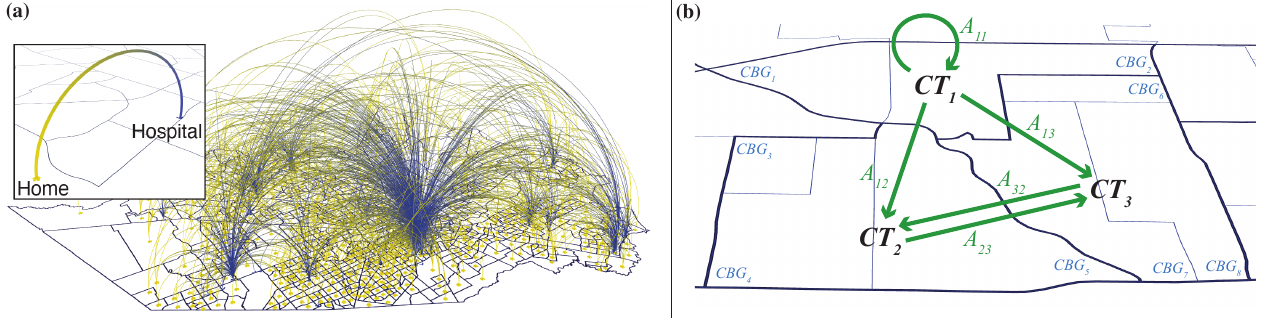}
\caption{\rev{{\bf Networks summarize human movements between census tracts.}}
%
%{\bf Example SafeGraph movement data and spatial aggregation across census tracts (\revrev{census tracts}).}
{\bf(a)} Visualization of observed movements from home neighborhoods to hospitals in Harris County, TX \rev{during the storm week beginning on February 15, 2021 (Monday).}
Home locations are recorded using U.S. census block groups, %$\mathcal{H} = \{CBG_i\}_{i=1}^H$, 
whereas destinations locations are Points of Interests (POIs)
%$\mathcal{P}=\{POI_i\}_{i=1}^P$  
with known latitudes, longitudes, and other information such as industry category (e.g. hospitals). 
%The movements are naturally represented by a weighted, bipartite network with a time-varying adjacency matrix $B(t)\in\mathbb{R}^{H \times P}$.
%
{\bf(b)} %In this work, we study a coarser network model that encodes movements between pairs of census tracts (\revrev{census tracts}), each of which consists of several CBGS. Letting $C$ denote the number of \revrev{census tracts} in Harris County, the resulting 
\rev{For different industry categories, we construct networks that are each encoded by a} time-varying adjacency matrix  in which $A_{ij}(t)$ encodes movements from home \revrev{census block groups spatially contained in a census tract, which we enumerate by $CBG_i$ and $CT_i$,} to POIs in \rev{census tract} $CT_j$ during week $t$. \rev{Much of our  study  focuses on studying the movements in and out of \revrev{census tracts} each week as defined by their node degrees: $d^{in}_i(t) =\sum_j A_{ji}(t)$ and $d^{out}_i (t) = \sum_j A_{ij}(t)$.}
}
\label{fig:1}
\end{figure}

In this work, we propose to study multilayer  mobility networks in which different layers  are defined according to the types of locations that persons visit---that is, the industry sector to which each location belongs. Different network layers are used, for example, to encode human movements to schools, grocery stores, hospitals, and so on. Our methodology involves studying observed movements using a cell-phone GPS dataset called SafeGraph \cite{SafeGraph2021} and constructing multilayer  networks that encode directed weekly flows between spatial regions. \rev{See Figure~\ref{fig:1} for an example illustrating  observed human movements from home neighborhoods to hospitals for Harris County, TX during the week of a 2021 winter storm.}
 Each network layer corresponds to an industry sector defined using the North American Industry Classification System (NAICS), which is a hierarchical classification scheme that gives rise to a hierarchy of network layers. %
This %modeling 
framework thereby allows for a rich, nuanced characterization, or ``fingerprinting'', for   human movements and movement changes and adaptations by industry sector that can occur, for example, seasonally or during disruptive events such as natural disasters. 
To illustrate this application, we apply this modeling framework to investigate how human mobility adapted during a winter storm.
By studying how people adapt their movement patterns with respect to different categories of movement  (e.g., visitations to schools,   hospitals, and grocery stores), our approach 
\rev{ examines ongoing and interrupted \emph{local movements}. 
This provides complementary insights to prior research on different risk-aversion behaviors such as sheltering at home \cite{gao2020mapping,coleman2020anatomy} and large-scale evacuations \cite{deng2021high,li2024using}---that latter of which is common for some disasters (e.g., earthquakes, hurricanes, and floods) but not winter storms \cite{wang2017aggregated}.}

%The objective of this study is to model human movement through a time-series of stratified networks, examine the impact of Winter Storm Uri on different movement categories, and look at correlations between movement direction and various factors including different population groups and infrastructure. 

%We expand on this work by developing time-varying, multilayer network models to study weekly movements between spatial regions, i.e., the census tracts within in Harris County.
%
%For each week, we construct multilayer networks (specifically, multiplex networks) in which each layer encodes a specific category of movement (e.g., health care, schools), where each . This framework of time-varying, multilayer networks allows us to analyze both category-specific behaviors and changes over time. 

Herein, we focus on human mobility adaptation during the 2021 Texas winter storm, or Winter Storm Uri, which hit Texas during February 13-17, 2021 and led to 246 deaths and more than \$195 billion damages \cite{winterUri2022}. %causing one of the state's worst natural disasters with 246 deaths and 9\% of those being caused by motor vehicle accidents \cite{mcullough_texas_winter_storm_2022}. 
This extreme weather event caused a disruption in typical human mobility patterns due to poor road conditions \cite{mcullough_texas_winter_storm_2022}, the inability of people to leave their homes, government recommendations to stay home \cite{mcentire, ReadyHarris2021WinterFreeze}, \revrev{and building closures} \cite{Click2Houston2021Closures}. There was also a huge impact on key infrastructure, including water and power outages. %, creating a need to divert certain activities like dialysis care from regular offices to hospitals \cite{cnn_hospital}. 
Previous research on this event has focused on the state's infrastructure including the power grid \cite{ZHOU2024104339}, water infrastructure resilience \cite{TIEDMANN2023104417}, and social disparities during outages in these systems \cite{grineski2023social}. Other studies have used cell phone location data to examine the  disproportionate impacts of this winter storm on different socioeconomic groups and community resilience \cite{lee2022community,chen2023enhancing}.

Complementing these studies, our utilization of multilayer mobility networks provides a fine-grained characterization of the impacts of Winter Storm Uri on human movements to locations associated with different industry sectors. We first investigated which layers of the network were the most / least impacted by the storm, finding that people largely reduced their movements to ambulatory healthcare services, restaurants, and schools, but prioritized movements to grocery stores and gas stations.
Much of our work focuses on understanding the network layers' in- and out-degrees that encode the cumulative movements into and outward from census tracts (defined according to each industry sector). We integrate additional data from the U.S. Census, including demographic, socioeconomic, and infrastructure information, and train models for in- and out-degree predictions during the storm week and other weeks. We find that in-degrees are  generally harder to predict than out-degrees, complementing known insights about the predictability of human movements \cite{song2010limits,yang2014limits,cuttone2018understanding}.  Interestingly, the predictability of out-degrees was not significantly impacted by the storm (with an R-squared score reduction of less than 1\%), while the predictability of in-degrees decreased significantly during the storm week (with an R-squared score reduction of 4-13\%).

Our work contributes to human behavior research during catastrophic events, aiming  to obtain a deeper understanding of people's adaptation and resilience to natural disasters by industry sector. Specifically, our work provides insights into which types of human movements are  prioritized (e.g., those related to basic needs such as food, water, and shelter) and which are strategically reduced. Our findings about the predictability of movements into and out of census tracts can also aid emergency planning and  disaster management for future extreme weather events.
In short, our approach of using multilayer mobility networks  to study the reduction, prioritization, and predictability of human movements categorized by industry sector broadens the understandings of how people adapt their mobility during  situations of heightened risk.
%

%\sout{
%This paper is organized as follows. In Section \ref{sec:methods}, we present our methodology  for developing multilayer networks in which layers encode movement categories defined according to industry sectors.
%%
%In Section~\ref{sec:results}, we present our main findings that characterize the storm's impact on network connectivity including the node degrees which encode  cumulative flows into and out of census tracts.
%%
%Finally, we discuss and summarize this work in Section \ref{sec:discussion}. 
%}

\rev{
This paper is organized as follows. In Section \ref{sec:results}, we introduce our main results which include methodology to construct multilayer mobility networks 
%with layers that are associated with different
% encode movement categories defined according to 
%industry sectors 
(Section~\ref{sec:network_creation})
%--\ref{sec:strat}). 
%Subsequently, we 
and investigations into the storm's impact on the 
reduction / prioritization of, and predictability of sector-specific movements (Sections \ref{sec:impact}--\ref{sec:census}).
%
%We then present our main findings that characterize the storm's impact on network connectivity including the node degrees which encode  cumulative flows into and out of census tracts.
%
In Section, \ref{sec:discussion} we discuss and summarize this work. 
Methods are are presented in Section~\ref{sec:methods}.
}

% Figure 1: small network illustration

%------------------------------------------------------------------------------------------------------------------------%------------------------------------------------------------------------------------------------------------------------
\section{Results}\label{sec:results}
%------------------------------------------------------------------------------------------------------------------------%------------------------------------------------------------------------------------------------------------------------
%

% Figure 2: network stratification by industry
\begin{figure}[t!]
\centering
\includegraphics[width=\linewidth]{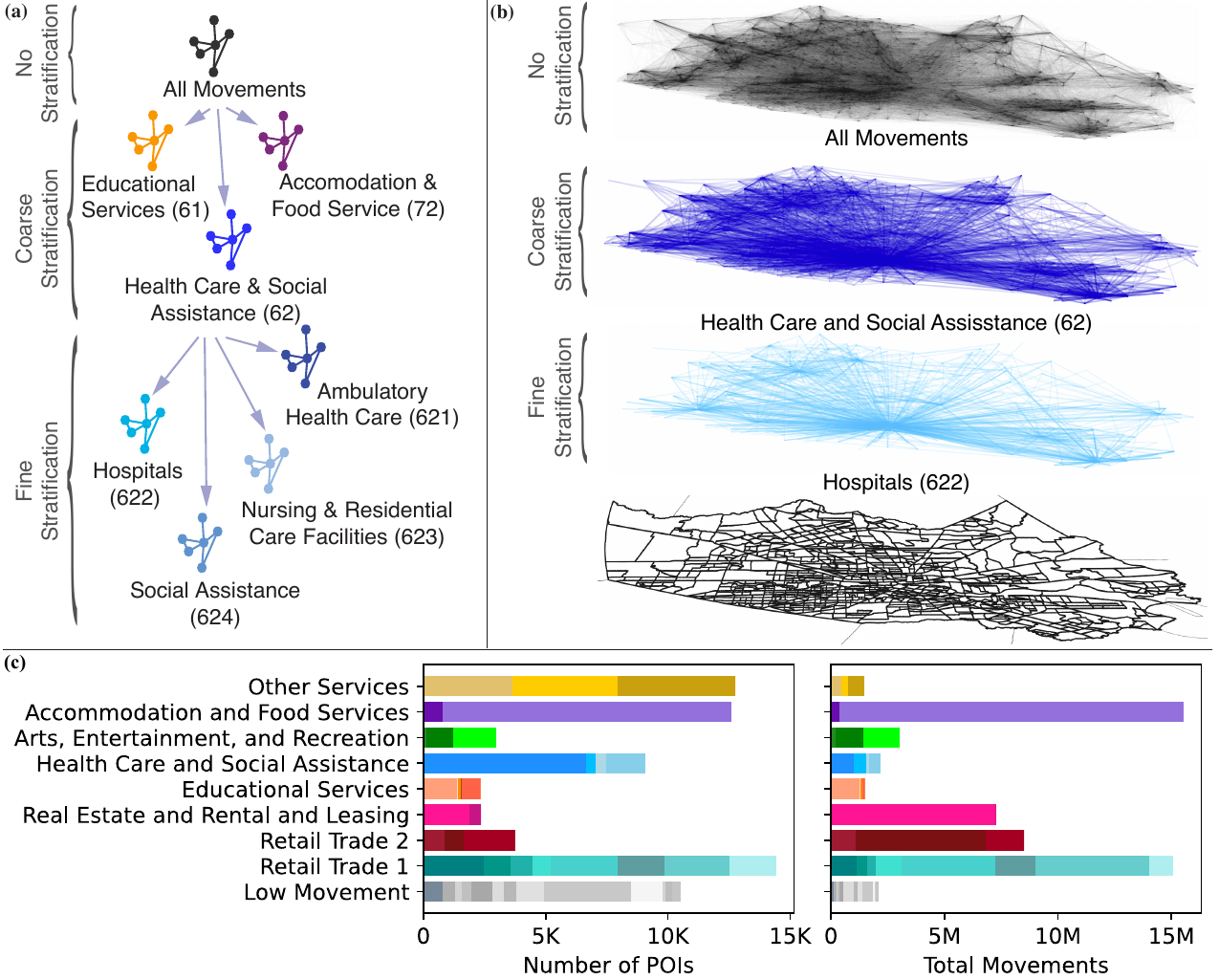}
\caption{\rev{{\bf Hierarchical stratification of movement categories by industry sector.}}
{\bf(a)} 
Toy illustration for the hierarchical stratification of a mobility network into network layers that encode different behavioral categories of movements, \revrev{defined using the North American Industry Classification System (NAICS).} The number of digits in a NAICS code determines the hierarchy depth (i.e., level of coarseness when refining movements categories into subcategories).
{\bf(b)} 
A map of \revrev{census tracts} in Harris County (bottom) overlaid by three example networks at three different coarseness levels: all movements (top), health care and social assistance (middle), and hospitals  (lower).
{\bf(c)} Fraction of POIs in each NAICS category for Harris County (left) and fraction of total observed movements in each NAICS category (right). 
%
%Total movements were summed over the entire timeframe per category, see Equation \ref{eq:movement}.  
In both \revrev{panels, for each category different coloration indicates finer subcategories.}
% charts, the outer rings show the categorical stratification according to a coarse scale with 2-digit NAICS codes, whereas the inner rings show  a finer stratification using   NAICS codes either three or four digits.
%Categories with very low movement are grouped together (grey). 
\rev{See Figure \ref{fig:4}  and Supplementary Figure 1 for additional details about the stratification of categories into subcategories and their industry sector NAICS codes.}
}
\label{fig:2}
\end{figure}

%===================================================
%{\color{revisions}
\subsection{
%\sout{Construction of a network that encodes movements between census tracts (\revrev{census tracts})}
%\\
\rev{Multilayer networks encode human movements to different industry sectors
%Networks models for movements between census tracts (\revrev{census tracts})}
}
\label{sec:network_creation}
}
%===================================================

\rev{To develop a nuanced characterization of the storm's impact on different categories of human movement, we first introduce a modeling framework involving time-varying, multilayer networks. Different  layers in the multilayer network represent observed movements  associated with different industry sectors. 
Our study area is Harris County, TX, which was severely affected by the 2021 winter storm,   and the study duration is 25 weeks beginning on Monday December 28, 2020 and ending on Sunday June 27, 2021. We enumerate these weeks $t=1,\dots,25$ and note that the storm's most severe impacts occurred on February 15-17 during week 8.
Following the literature \cite{nejat2022equitable,xu2023power,lee2022community}, we aggregate \revrev{census block groups} and POIs across census tracts to yield networks that summarize observed movements from one \revrev{census tract} to another, and   the movement destinations are associated with a particular industry category (e.g., hospitals).
%construct the network based on \revrev{census tracts} and POIs.
See Figure \ref{fig:1} for a visualization and Section~\ref{sec:construct} for further details.}

\rev{We construct different networks for different industry sectors, and} 
we refer to the act of \rev{separating} a network's edges   into categorized sets of edges associated with network layers as ``stratification''  \cite{stanley2016clustering}.
 %We construct behavior-stratified multilayer networks in which  different network layers encode different behavioral categories of movement (e.g., visits to schools, hospitals, etc.).
 More specifically, we study a type of multilayer network called a multiplex network in which each layer consists of the same set of nodes (i.e., in our case, the set of \revrev{census tracts})
 \rev{and in our case each layer encodes different behavioral categories of movement (e.g., visits to schools, hospitals, etc.).} 
 We classify behavioral categories of movement based on the 2017 North American Industry Classification System (NAICS), which were used to classify  POIs in the SafeGraph data. 
Importantly,   NAICS   is a hierarchical categorization scheme, allowing us to stratify movement data into a hierarchical set of mobility network layers. 
\rev{Each NAICS category has a numerical code with 2 to 6 digits depending on level in the hierarchy, with two digits at the coarsest level and six digits at the finest, most-granular level. 
See Figure \ref{fig:2}(a) for a toy illustration of this hierarchy of network layers.
For each NAICS code $n$, we define a time-varying  adjacency matrix so that $A^{(n)}(t)$ describes network layer $n$ during week $t$.  %For example, Figure \ref{fig:1} depicts $A^{(622)}(8)$, i.e., the network for hospitals (NAICS code 622) during the storm week ($t=8$). 
(See Section \ref{sec:construct} for further details.)}
%
%That is, at a coarse level of the hierarchy, the network encoding all movements can be stratified into into layers encoding movements to Educational Services (NAICS code 61), Health Care and Social Services (62), Accommodation and Food Service (72), and additional categories. Whereas at a finer level of the hierarchy, the network layer encoding movements to Health Care and Social Services can be further stratified into  network ``sublayers'' that encode, e.g., movements to locations associated with Ambulatory Health Care Facilities  (621),  Hospitals (622), and so on.
 %
In Figure \ref{fig:2}(b), we depict  a map of \revrev{census tracts} in Harris County TX overlaid with visualizations of example network layers at 3 different levels of coarseness for the movement categories: all movements (top), movements to health care and social assistance location (middle), and movements to hospitals (lower). 
%
%In our study, the three-digit layer provides a sufficient breakdown for fine stratification, except for educational services where we consider four digits of NAICS codes. Each NAICS category has a numerical code with 2 to 6 digits depending on level in the hierarchy, with two digits at the coarsest level and six digits at the finest, most-granular level.  At the coarsest level of the hierarchy, each NAICS category has a 2-digit numerical code indicating whether the POI is associated with one of twenty-four  industry classifications.

\rev{Our study will examine multilayer networks with layers associated with two hierarchy levels categories. At the coarsest level, we focus on the eight industry categories with the highest movement:  retail trade 1 (NAICS 44); retail trade 2 (45); real estate and rental and leasing (53); educational services (61); health care and social assistance (61); arts, entertainment, and recreation (71);  accommodation and food services (72); and  other services (81). See Section \ref{sec:threshold} for details on how we chose which categories to focus on. For the finer hierarchical level, we identified industry sectors according to the three-digit NAICS codes; however,  for educational services (61) we used four-digit codes since the categories for 61 and 611 are identical.
}

\begin{figure}[b!]
\centering
\includegraphics[width=\linewidth]{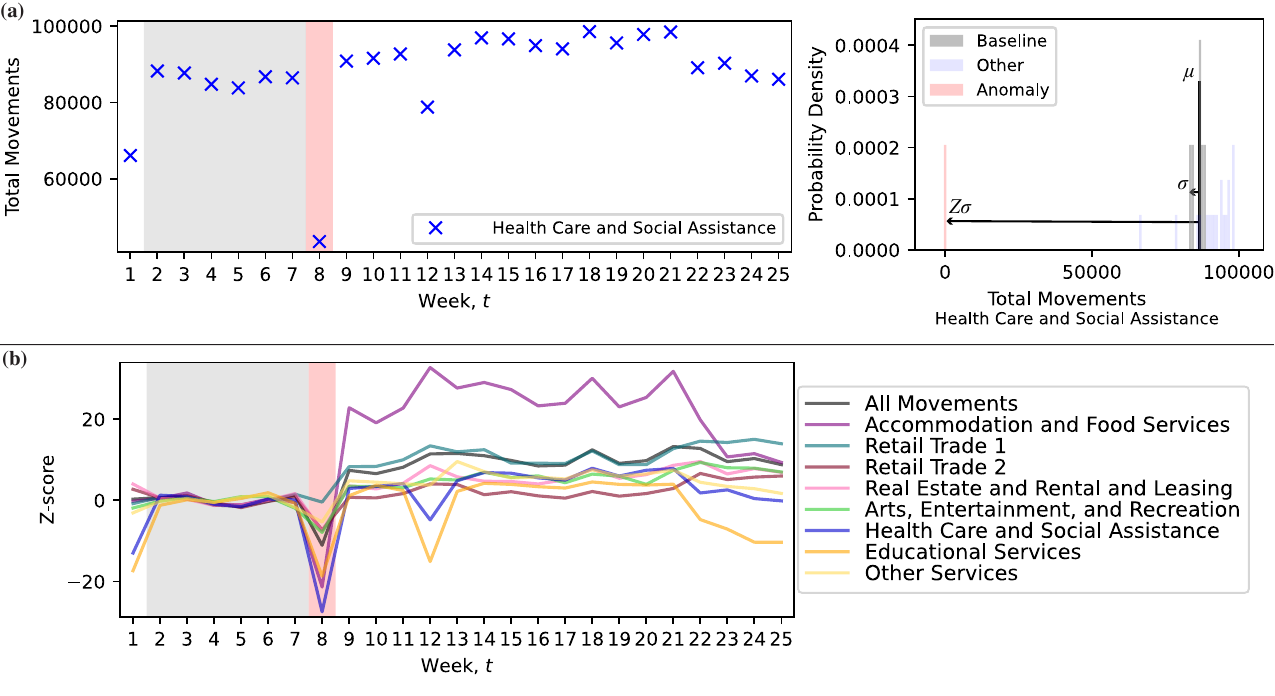}
\caption{
%\sout{{\bf Decreased movement during storm across industry categories.}}
\rev{{\bf Decreased movement during storm, quantified by z-scores.}}
%Z-scores quantify impact of storm on eight largest movement categories.}
{\bf (a)} 
\rev{We plot the total movements  $m^{(n)}(t)\rev{=\sum_{i,j} A_{ij}^{(n)}(t)}$ during each week $t$} for the network layer \rev{that encodes observed} movements to locations associated with health care and social assistance \revrev{(NAICS code 62)}. Red and gray shading highlight the storm week and the weeks  
%$\mathcal{T}_{base}$ 
used to construct a baseline \rev{for comparison}.
\rev{We quantify the change in movements during the storm week, $t=8$, using a z-score $  Z^{(n)}(t) \approx -27$, which is visualized in the right-hand panel and is discussed in Section~\ref{sec:zscore}. It's calculation  uses a baseline mean, $\mu^{(n)}$,  and standard deviation, $  \sigma^{(n)}$.
{\bf (b)}
We plot the z-scores, $Z^{(n)}(t)$,  versus $t$ for the eight NAICS categories with largest total movements across the 25-week study duration.  }
\vspace{-.5cm}
%(e.g., $M^{(n)} > 10^{6}$)
%Weeks $\{2, 3, \dots, 7\}$ as the baseline (grey) and week 8 as the storm week (red). 
%Note that the movement categories are ordered from most-to-least impact by the storm (week 8), and the NAICS  descriptions have been shortened (e.g. `Accommodation and Food Services' $\mapsto$ `Accommodation').   Green and yellow shading highlights Z-scores greater than 2 and less than -2, respectively. 
}
\label{fig:3}
\end{figure}

%===================================================
\subsection{
%\sout{Identifying the most and least impacted movement categories}
%\\
\rev{Movements significantly decreased during the storm week}
}\label{sec:impact}
%===================================================

%\sout{
%Following the methodology introduced in Sections~\ref{sec:network_creation} and \ref{sec:strat},
%we constructed multilayer networks with weighted, directed edges that encode movements
%among census tracts, using  network layers to encode different behavioral categories of movement (e.g., visits to schools, hospitals, etc.). Each layer is indexed by a NAICS code $n$ that contains 2--6 digits depending on the hierarchy level that is chosen to classify locations (and categorize movements to them).
%}
%

\rev{Using multilayer networks encoding high-movement industry categories, we study their structure to investigate the impact of the storm on human behavior. Our approach relies on statistical analyses of the node degrees for the network layers and the ``aggregated network'' that does not distinguish movement categories.}
%
%\rev{Using the network modeling introduced in Sections~\ref{sec:network_creation} and \ref{sec:strat}, 
%we next quantify the storm's impact using statistical methods.}
%
Beginning with the coarsest level of movement categorization (i.e., 2 digit NAICS codes), for each $n$ we examined the time series   
\rev{$m^{(n)}(t) = \sum_{i,j} A_{ij}^{(n)}(t)$}
of total movements for a 25-week study duration from December 29, 2021 to June 28, 2021 and computed z-scores $Z^{(n)}(t)$  (see Section~\ref{sec:zscore}) to identify statistically significant differences between $m^{(n)}(t)$ and baseline    values that were found using the six weeks preceding the storm.
%

%We chose to display movement categories with the largest total movements, $M^{(n)}>10^{6}$, since these categories make up the majority of movements in Harris County. Note that NAICS descriptions have been shortened. Z-scores are calculated using Equation \ref{eq:zscore} and are shaded in green if greater than 2 (indicating an increase in movement) and in yellow if less than -2 (indicating a decrease in movement). We can now look at the storm's impact on the stratified networks. The z-scores are calculated using baseline (grey), $\mathcal{T}_{base} = \{2, 3, \dots, 7\}$, and are sorted by decreasing magnitude for the storm week (red), $Z^{(n)}(8)$. 

% table: z-scores
%\begin{table}[htpb!]
%\centering
%\includegraphics[width=\linewidth, trim=0 600 0 50, clip]{figs/zscoretable.pdf}
%\caption{{\bf Z-scores quantify impact of storm on eight largest movement categories.}
%For the eight NAICS categories $n$ with largest total movements, we compute z-scores $Z^{(n)}(t)$ (see  Section~\ref{sec:zscore}) for the 25-week study duration.  
%%(e.g., $M^{(n)} > 10^{6}$)
%%Weeks $\{2, 3, \dots, 7\}$ as the baseline (grey) and week 8 as the storm week (red). 
%Note that the movement categories are ordered from most-to-least impact by the storm (week 8), and the NAICS  descriptions have been shortened (e.g. `Accommodation and Food Services' $\mapsto$ `Accommodation').   Green and yellow shading highlights Z-scores greater than 2 and less than -2, respectively. 
%}
%\label{tab:zscores}
%\end{table}

\rev{A visualization of this calculation is provided in  Figure \ref{fig:3}(a) for an example
network layer encoding movements to locations associated with Healthcare and Social Assistance (i.e., NAICS code 62). We find
$Z^{(62)}(8) \approx -27$, implying that these movements %encoding in this network layer is tremendously decreased during the storm, i.e.,
significant decreased during the storm, i.e., by approximately 27 standard deviations.}
\rev{In Figure~\ref{fig:3}(b), we plot z-scores $Z^{(n)}(t)$ for the high-movement categories across the 25-week study duration. }
%Each line represents a  movement category. We also emphasize that we've focused here on the eight movement categories having the most observed movement. (Recall the pie chart in Figure~\ref{fig:2}(c) that depicts the fraction of observed movements for each category.)
%
Observe %in Figure~\ref{fig:3}(b) 
that all of the coarse-level categories of movement that we considered exhibited a decrease   during the storm week ($t=8$). %(Recall that we focus here on the categories associated with most movements, as illustrated in Figure~\ref{fig:pies}.)
%Further note that we've ordered the table's rows from most-to-least impacted movement categories. 
The most impacted movement categories are health care and social assistance (62), accommodation and food services (72), and educational services (61).
%
%%
% interpreting zscores
%The z-scores highlight the three movement categories most impacted by the storm: health care and social assistance, accommodation and food services, and educational services. 
Movements in these categories are significantly reduced, which is likely due to the closures of hospitals, schools, and restaurants during the storm. In contrast, retail trade 1 (44), which includes grocery stores and other essential food vendors, appears to have been the least affected. Movements to these locations were prioritized despite the heightened risk imposed by the storm. 
For the weeks following the storm,  movements increased across all categories, which is aligned with a seasonal trend that occurs each spring. See Supplementary Figure 2 for multi-year time series showing this trend across movement categories.

\rev{Before continuing, we highlight that the storm's impact appears to occur exclusively during week 8 (February 15-21, 2021), which is expected since the most severe effects (e.g., blackouts and deaths) occurred on February 15-17. It's worth pointing out that the network data is aggregated across a larger time window (i.e., the full week) but the anomalous storm largely caused network structural changes primarily during a subset of those days.  Aggregating temporal network data across a larger time window is known to cause network properties to have a diminished signal strengths \cite{caceres2011temporal, taylor2017super}. Here, we expect that the z-scores would generally increase (i.e., enhanced signal detection) if we were able to select a time window to perfectly align with the storm days. However, the dataset we study is provided at the weekly timescale; nevertheless the anomaly signal is very strong.}

\rev{That said,} there are several other anomalous decreases in movement for some categories. Week 1 includes the holiday of New Years, and we observe that this week   has decreased movement to locations associated with health care (62), education (61) and other services (81) but increased movement to locations associated with real estate (53) and retail trade 2 (45).
In addition, decreased movements to health care and education facilities occur during week 12, which we predict occurs due to the school closures and increased vacationing that occurs during spring break.
%
%compared to the baseline week (weeks 2-6), which 
%We can see other weeks where movement is anomalous in some categories to the baseline. Specifically, a decrease in movements during Week 1 and Week 12 for health care and educational services, we attribute this to winter and spring breaks. 
Finally, starting week 22 we observe decreased movement to educational facilities, which likely occurs due to the start of summer break. 

 \rev{
We also note that our baseline weeks coincide with the end of the COVID-19 period \cite{limon2020coronavirus} and acknowledge the difficulty to disentangle the lingering effects of the pandemic from the storm. In Supplementary Figure 2, we see for our study duration that the majority of movement categories had returned to their pre-pandemic numbers, with the exception of educational services, health care, and accommodation (which could be considered the new normal movement patterns). 
}

\subsection{\rev{Storm impact on movements  with a finer stratification of industry sectors}
}\label{sec:finer}

% explaining sankey
So far, we have only considered movement categories (i.e., network layers) defined at a coarse scale in which the mobility network is stratified into coarsely defined movement categories using 2-digit NAICS codes. However, 
%as discussed in Section~\ref{sec:strat}, 
NAICS is a hierarchical classification scheme allowing us to stratify movement categories (and their associated network layers) into a hierarchy. Next, we extend our study of z-scores by  considering a finer stratification of movement categories using 3-digit NAICS codes (except for educational services for which we used 4 digits, since using 3 digits does not provide a finer stratification.)
%To better understand the storm's impact we further examined a finer stratification of the movement categories.

\begin{figure}[h!]
\centering
\includegraphics[width=\linewidth]{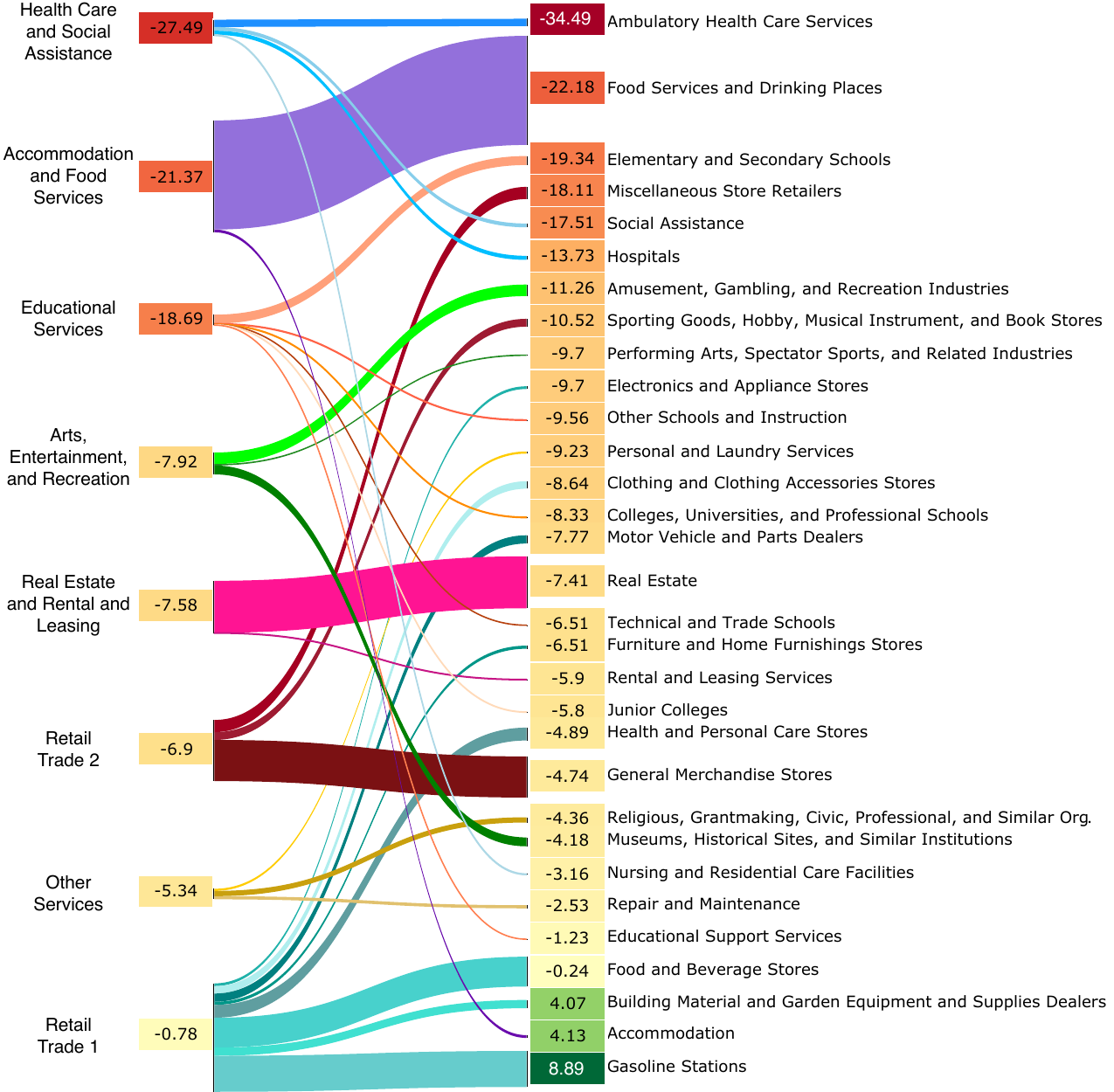}
\caption{{\bf Comparing the storm's impact on movement categories and sub-categories.}
Z-scores quantify the storm's impact on  movement categories defined using the NAICS hierarchical classification scheme. These are shown using both a coarse scale with 2-digit NAICS codes (left) and a finer scale using 3 or 4-digit NAICS codes (right).
\rev{See Supplementary Figure 1 for the industry sector NAICS codes.}
Both sets of movement categories  are ordered top-to-bottom based on their computed z-scores \revrev{(shown in colored boxes)} so the most-decreased movement categories are at the top. Curved lines depict how coarse movement categories separate into  finer categories, and the line widths are proportional to the number of observed movements for each category.
%Node thickness represents the number of POIs in Harris County: health care (9066 POIs), accommodation (12590 POIs), and educational services (2323 POIs). Category 6114 removed due to low visits during all weeks. 
}
\label{fig:4}
\end{figure}

In Figure \ref{fig:4}, we visualize the z-scores during the storm week for a coarse stratification of movement categories on the left and a finer stratification  on the right. Both sets of NAICS codes (i.e., coarse versus fine) are ordered from top-to-bottom in order of z-score so that the most decreased movement categories are at the top. Curved lines show how each coarse movement category separates into finer categories, and the line widths are proportional to the  total movement for each category. We also note that the category \revrev{containing business schools and computer and management training (6114)} is omitted due to the observed movements being too small (i.e., only 2  were observed).

%number of observed POIs in Harris County for each category: health care and social assistance has 9066 POIs with the majority from 6666 ambulatory (non-hospital) health services, accommodation has 12590 POIs with 11822 food services and drinking places, and educational services has 2323 POIs with 1372 elementary and secondary schools. 
%We use the level of NAICS hierarchy that best shows the breakdown for the coarse category, a three digit level for health care and accommodation categories and four digits for education.  
%We have removed 6114 Business Schools and Computer and Management Training, because there were only a few visits during 2 weeks. The finer hierarchy in NAICS classification allows a more in-depth view of industries affected by the storm.

% interpreting sankey
We first highlight that there is remarkable consistency between the 3 most-impacted movement categories at the coarse scale and at the fine scale. The 3 most-impacted coarse movement categories were (62) healthcare, (72) accommodation and food services, and (61) education.
At the finer scale, the 3 most-impacted subcategories are derived from these 3 categories, one each, and their z-scores retain the same order.
The most impacted fine-scale movement category is ambulatory heath services, $Z^{(621)}(8) = -34.49$, which includes POIs such as physician and dentist offices, outpatient care centers, and home health care services. The second is food service and drinking places, $Z^{(722)}(8) = -22.18$, and further examination revealed that restaurants is most impacted sub-sub-category (i.e., $Z^{(7225)}(8) = -21.37$). (We must note that the COVID-19 restriction on restaurant capacity in Harris County had been at 50\% during the storm week and was only raised back to 100\% on March 10, 2021 \cite{opentexas}.) Lastly, movements to elementary schools is the third most-impacted fine-scale category, $Z^{(6111)}(8) = -19.34$, while other educational institutes like universities and junior colleges were less impacted.

Importantly, Figure \ref{fig:4} also reveals which categories of movement were prioritized during the storm. Movements to food and beverage stores (445) decreased very little, and at the same time, movements actually increased to three types of locations:
gasoline stations, $Z^{(447)}(8) = 8.89$ (which are  critical infrastructure and offer easily accessible food),
accommodations, $Z^{(721)}(8) = 4.13$ (which includes hotels for dislocated peoples but also has a regular seasonal increase shown in Supplementary Figure 2), and
building materials, $Z^{(444)}(8) = 4.07$ (which includes home stores including Home Depot and Lowes).
%We see an increase in visits to accommodation services, $Z^{(721)}(8) = 4.13$, this category includes hotels and motels. Previous years show a general increase in this category during this time of year. See Appendix \ref{apx:movements} for total movements during previous years and a finer resolution for high-movement categories. Total movements given insight into which movement categories were most affected by the storm. Next we look at the direction of the movements, by looking at out-degree and in-degree.
%

% Figure 4: zscores sankey

%===================================================
\subsection{Storm's impact  on mobility networks' in- and out-degrees}\label{sec:mixing}
%===================================================

\begin{figure}[t!]
\centering
\includegraphics[width=.95\linewidth]{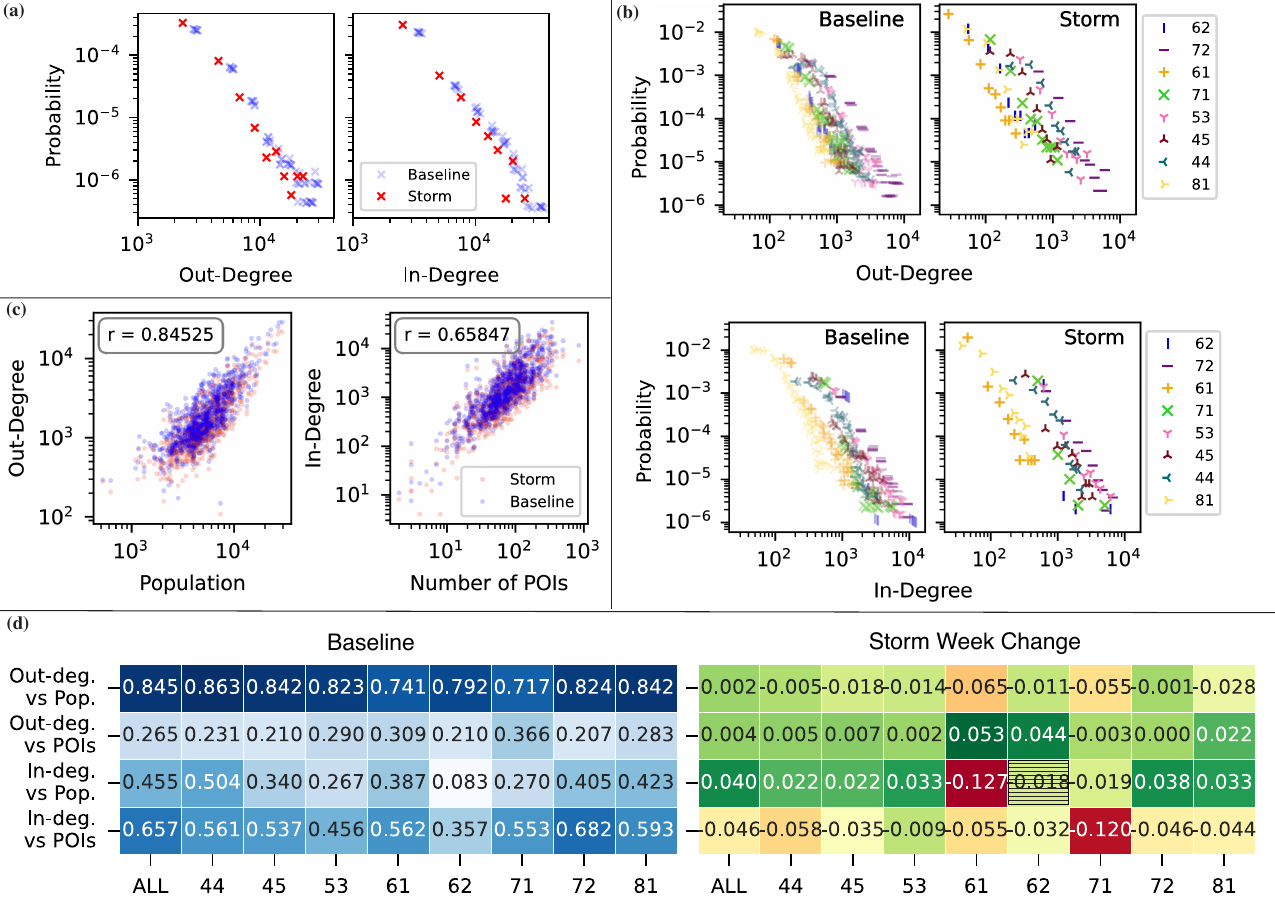}
\caption{{\bf Node degrees reveal heterogeneous flows among \revrev{census tracts}.}
% and their association populations and POI counts.}
{\bf(a)} \rev{We show distributions} of in-degrees $d^{in}_i(t)$ and out-degrees $d^{out}_i(t)$  for the mobility network combining all movement categories (left) during the six baseline weeks (blue) and  storm week (red). 
{\bf(b)}
Focusing on network layers associated with high-movement categories, we plot the distributions of in- and out-degrees for both the baseline weeks and the storm week. 
The probabilities  decay linearly in a log-log scale suggesting a power-law relation.
%, highlighting broad heterogeneity.
%degree observations plotted for each week. Symbols (x) represent the likelihood a random \revrev{census tract} will have an approximate degree for a given week, 
{\bf(c)} Scatter plots reveal that (left) a \revrev{census tract}'s out-degree is strongly correlated with the population residing in that \revrev{census tract}, and (right) a \revrev{census tract}'s in-degrees is strongly correlated with its infrastructure (i.e., the number of POIs in the \revrev{census tract}).
%with their population sizesRelationship between POIs and in-degree, and population and out-degree across \revrev{census tracts}. Each symbols (o) represents a \revrev{census tract}'s degree for either the mean across baseline weeks (blue) or the storm week (red). 
%(Pearson r-values are indicated.)
% and p-values are much less than 0.05.
{\bf(d)} For high-movement categories during the baseline weeks,  Pearson correlation coefficients ($r$-values)
measure correlation between \revrev{census tracts}' in- and out-degrees versus their populations and the number of POIs for each industry sector. 
%Shading indicates the strength of the correlation, with darker green representing higher r-values. 
%
We report how the  r-values changed during the storm week (i.e., $r$ for storm week minus $s$ for the baseline weeks).  \rev{(We note that all p-values were smaller than 0.05 except for one instance, which is outlined by a black box.)}
}
\label{fig:5}
\end{figure}

%To further understand how movement patterns   changed during the 2021 winter storm, w
In this section, we study the in-degree $d^{in}_j(t)$ and out-degree $d^{out}_j(t)$ that encodes the weekly movements into and out of, respectively, each census tract $CT_j$. We note that  we study ``weighted degrees'' (which are also commonly called node ``strengths'').
%, considering both networks that encode all movement types as well as network layers that encode movements  within particular NAICS categories.
%
%First, we study changes in the distributions of in-degree and out-degrees, capturing how overall connectivity was affected. 
%
In Figure \ref{fig:5}(a), we show  distributions of in- and out-degrees during the storm week (red) and during the six baseline weeks preceding the storm (blue). These \rev{distributions} were computed across the 786 \revrev{census tracts} in Harris County using 10 bins.
Observe that both degree distributions  appear linear in a log-log scale, which suggests a power-law relation (although there is limited evidence, since the degree heterogeneity spans only about 1.5 decades). Because network connectivity decreases during the storm, the node degrees decrease during the storm, which manifests as a shift-left for the degree distributions. Interestingly, the degree distributions do not otherwise significantly change. 
In Figure \ref{fig:5}(b), we show that similar degree distributions arise for   the network layers that encode different movement categories, and they are similarly impacted by the storm.
% (with the largest shifts left occurring for the most impacted movement categories).

%We examine the relationship between the number of POIs and in-degree, as well as between population and out-degree across \revrev{census tracts}. 
To help understand the origin (or main drivers) of   degree heterogeneity across \revrev{census tracts},
% inward and outward movements,  
next we support two hypotheses: 
\emph{\revrev{census tracts} with large (or small) populations should have many (or few) outward movements; and \revrev{census tracts} containing many (or few) POIs should have many (or few) inward movements.}
Thus motivated, in Figure \ref{fig:5}(c) we  plot (left) $d^{out}_i(t)$ versus \revrev{census tract} population size and (right) $d^{in}_i(t)$ versus the number of POIs, respectively, for the \revrev{census tracts} in  Harris County. Both pairs of variables exhibit significant correlation with Pearson correlation coefficients
%, as indicated in the figures. 
%
%the number of POIs versus in-degree and population versus out-degree, each symbol (o) represents a \revrev{census tract}, with blue symbols denoting the mean values taken across the baseline weeks, and the red symbol representing the storm week. 
%To assess the strength and direction these correlations, we calculate the Pearson correlation coefficient (r-value) for both plots. The Pearson r-value for the relationship between POIs and in-degree is 
given by $r\approx  0.85$   and $r\approx0.66$, respectively, 
\rev{with p-values within numerical precision of zero.}

%$b_1, \dots, b_{10}$, to represent the degree distribution across the 786 \revrev{census tracts}, note that each week has unique bin edges. 
%We estimate the probability that a randomly selected \revrev{census tract} has a degree falling within a given bin, $P(d \in b_i) = \frac{c_i}{786|b_i|}$, where $c_i$ is the count of observations in $b_i$ and $|b_i|$ is the size of $b_i$. Each symbol (x) represents the likelihood a \revrev{census tract} will have a given degree for a specific week. We observe for both out- and in-degrees, a lower likelihood as magnitude of degree increases, as well as a decrease in degrees. To gain more insight into the relationships between movement direction and community factors we perform correlation analyses.
%
%Note that each in-degrees $d^{in}_j(t)$ and out-degree $d^{out}_j(t)$ represents the observed weekly movement into and out of each census tract $CT_i$, respectively. (There are 786 \revrev{census tracts} in Harris County.)

% Figure 5: degree mixing

Next, we extend this correlation study to the network layers that encode different movement categories. That is, for each \rev{network layer associated with each} NAICS code $n$, we calculate each $CT_i$'s out-degree   $d^{out,(n)}_i(t)\rev{ \sum_{j} A^{(n)}_{ij}(t)}$ and in-degree $d^{in,(n)}_i(t)=\rev{ \sum_{j} A^{(n)}_{ji}(t)}$ and then calculate the associated Pearson correlation coefficients $r$ comparing these degrees to a \revrev{census tract}'s population and related infrastructure (i.e., the number POIs in that \revrev{census tract} having that particular NAICS code $n$). The associated r-values across baseline weeks are reported in Figure~\ref{fig:5}(d). For comparison, we also include correlations between out-degree vs. number of POIs and in-degree vs. population. In Figure~\ref{fig:5}(d) we show how each Pearson correlation coefficient changed during the storm week (right). Note that all correlations are statistically significant with p-values below $0.05$, except for the one value that is \rev{highlighted by a black box} (see NAICS 62 for the storm week).

Observe in Figure~\ref{fig:5}(d) that the strongest correlation occurs between out-degrees and \revrev{census tract} populations with $r\in[0.71,0.87]$ across all movement categories. The second-strongest correlation occurs between in-degrees and  the numbers of POIs in \revrev{census tracts}, with $r\in[0.35,0.69]$ across all movement categories.  We additionally observe correlation between out-degrees and POI numbers, and between in-degrees and population, however their associated r-values are generally smaller. Similar to our hypothesis for Figure~\ref{fig:5}(c), this suggests that even at the resolution of individual movement categories, population drives outbound movement, while local infrastructure attracts inbound visits. 
We also do not find much variation across different movement categories, with exception of movement category health care and social services (62), which has lower r-values for correlations relating to in-degrees. We predict this lower correlation occurs due to the nature of hospital infrastructure, i.e., fewer hospitals exist, and each serves as centralized hubs that attracts large numbers of visitors. 
Turning our attention to the storm week, we find that  the storm's effect on correlations is small.  The largest changes to $r$ occur for the correlation between in-degree and population for educational services (61)  and  between in-degree and POI numbers for arts and entertainment (71).
\subsection{Predictability of node degrees using demographic, socioeconomic, and infrastructure information}\label{sec:census}
%===================================================
%

%%Figure: census choropleths
%\begin{figure}[ht]
%\centering
%\includegraphics[width=\linewidth]{figs/choropleths.pdf}
%\caption{{\bf Socioeconomic and demographic data from U.S. Census.}
%Choropleth maps display   demographic and socioeconomic factors across \revrev{census tract} in Harris County, TX obtained from the U.S. Census. The maps show the distribution of population, income, percentage of non-white population, and poverty rate, offering a spatial overview of these key   characteristics. \revrev{census tract} 980000 is colored white, because the population is so small (i.e., 4 people). %We omit it from our regression analysis.
%}
%\label{fig:census}
%\end{figure}

In Section \ref{sec:mixing}, we supported our hypothesis that \revrev{census tract} population is a main driver for outward movements, while  POI infrastructure is a main driver for movements into \revrev{census tracts}. 
%We provided evidence for networks encoding either all movements or distinct categories, and we found the storm had little effect on these correlation.
%
We now use multivariate linear regression to perform a broader investigation for how network connectivity during normal times and the storm week are associated with demographic, socioeconomic, and infrastructure information.
That is, we obtain predictive models for \revrev{census tracts}' in- and out-degrees using infrastructure variables (i.e., the number of POIs in each \revrev{census tract}) and six social factors from U.S. Census data: population, population density, income, non-white percentage, poverty rate, and unemployment rate. See Section~\ref{sec:regression} for discussions on the dataset, this modeling framework, and our use of variance inflation factors to select a subset of social factors while preventing variable multicollinearity. To prevent multicollinearity for the infrastructure information, we  use either the total count of POIs across NAICS categories or separate counts for the   different NAICS categories. We restrict our models to the eight NAICS categories associated with the most movement (see Figure~\ref{fig:2}(c)).

\rev{In Figure \ref{fig:6}(a), we depict choropleth maps for Harris County that visualize two key features for the regression models: population and number of total POIs in each \revrev{census tract}. We additionally visualize the \revrev{census tract}'s mean out- and in-degrees across the baseline weeks. %Through the shading we see strong observable correlations between population and out-degree in certain areas, especially \revrev{census tracts} with high population. We also see correlations between number of POIs and in-degree (although not as obvious).}
In Figure \ref{fig:6}(b), we provide a visualization to illustrate our multivariate regression analysis that predicts \revrev{census tracts}' out-degrees using two social factors (population and income) and no infrastructure information. In the top \revrev{panel}, we illustrate the 2-dimensional regression plane (yellow) and the observed values across \revrev{census tracts} (blue). 
%In the top right we compare the \revrev{census tracts}' predicted and observed out-degrees. 
In the bottom, we quantify prediction accuracy using R-squared scores ($R^2$), which in this case is given by $R^2\approx 0.748 $. Note this prediction accuracy outperforms linear regression using just \revrev{census tract} population, since for single-variable regression the R-squared score is given by the square of the Pearson correlation coefficient: 
$R^2 \approx 0.84525^2\approx0.714 $. That is, including income (i.e., as well as population) yields a 4.7\% accuracy improvement for predicting movements outward from \revrev{census tracts}.
We provide bar graphs that summarize  $R^2$ and the  regression coefficients (left). Note   the coefficient for population is much larger than that for income, highlighting   population is  the more-important social factor for out-degree predictions. The  bar graph on the right indicates how   $R^2$ and the regression coefficients change if   the regression model is fit to data restricted to the storm week; $R^2$ changes minimally, but the coefficients decrease by 20-26\%. }

\rev{
In Figure \ref{fig:6}(c), we report $R^2$ and regression coefficients for   several regression models. In models 1 and 2, we study how movements outward from \revrev{census tracts} are related to social factors. That is, we constructed two regression models that predict out-degrees using either:  only population (model 1) or all six social factors (model 2). 
Comparing the two models, observe that including the five additional social factors increase $R^2$ by 9.85\% (i.e.,  from 0.714 to 0.784). Also note that the largest regression coefficients (in order) are population, population density,  non-white percentage, and income. That is, we find these to be the most important \revrev{census tract} variables for predicting movements outward from \revrev{census tracts}.
For models 3 and 4, we study how movements into \revrev{census tracts} are related to industry information, i.e., number of POIs. We predict in-degrees using either: total POI counts, while ignoring NAICS codes (model 3) or stratified POI counts for different NAICS codes (model 4). 
Comparing models, observe that $R^2$  increases 14.9\% when POI counts are calculated separately for the different NAICS codes, and the most important codes (in order) are 53, 45, 44 and 72. Interestingly, these are the four NAICS categories associated with the highest movements (recall Figure~\ref{fig:2}(c)).

\clearpage

%Figure 6: Regression
\begin{figure}[th!]
\centering
\includegraphics[width=\linewidth]{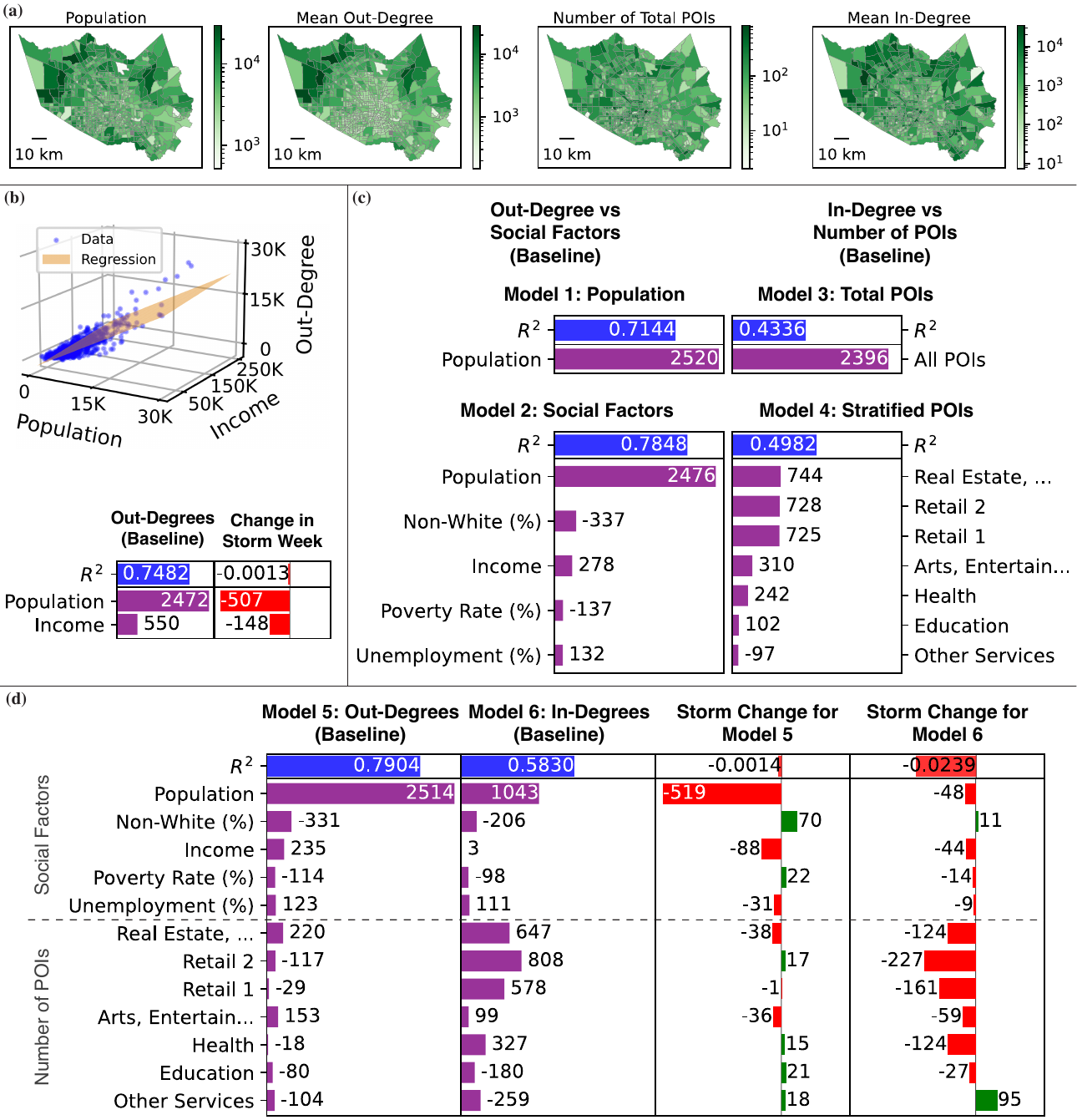}
\caption{
\rev{
{\bf Multivariate linear regression for predicting inward and outward movements}
{\bf(a)} 
Choropleth maps of \revrev{census tract}'s in Harris County, TX are used to visual two key regression features (populations and the number of total POIs in \revrev{census tracts}) and the two target variables (in- and out-degree). We note that  our analysis omits \revrev{census tract} 980000 for which the population and number of POIs is unusually small (i.e., 4 and 1, respectively), and we have colored that \revrev{census tract} gray.
% colored grey, because the population and the number of POIs are significantly lower than other \revrev{census tracts}.
{\bf(b)}
%~(Top left) 
Visualization of a multivariate linear regression model that predicts  out-degrees based on two input features (population and income) and takes the geometric form of a 2-dimensional plane that is fitted to empirical observations. (Each data point represents a \revrev{census tract} in Harris County.) 
We additionally provide 
%visuals for the comparison of observed and predicted out-degrees as well as 
the  model's R-squared score ($R^2$) that measure prediction accuracy as well as its regression coefficients. It is also indicated how the model changes when fit to data during the storm week.
% is also shown how these values change during the
%
%~(Top right) Comparison of observed and predicted values according to the regression model. 
%~(Bottom) Bar graphs are used to model performance (measured by an R-squared value) and the regression coefficients. Observe that the coefficient for population is larger than that for income, indicating \revrev{census tract} population is more important for predicting outward movements from \revrev{census tracts}. How the regression model changes during the storm week is shown in the right bar graph.
{\bf (c)}
We report $R^2$   and regression coefficients for four additional regression models that predict either out-degrees versus social factors  (models 1--2) or in-degrees versus number of POIs (models 3--4). 
%These models are fit to empirical data during the baseline weeks.
{\bf (d)}
Similar information is given for models 5 and 6 that use all input features. It is also indicated how these models change when fit to data during the storm week.
%
 %for both the baseline (left) and how $R^2$ and 
% regression mcoefficients changed during the storm week (right).
}
}
\label{fig:6}
\end{figure}

\clearpage

In Figure \ref{fig:6}(d), we study how movements outward and inward to \revrev{census tracts} are related to all features (social factors and industry information) for the baseline weeks (models 5 and 6)  and the storm week.
Comparing model 5 (predicting out-degrees using all features) to   model 2, observe that $R^2$ has a very modest 0.71\% increase and that the regression coefficients associated with POI counts are relatively small. Similarly, comparing model 6 (predicting in-degrees using all features) to the model 4, observe that  $R^2$ increases by a significant 17\%, and population has a very large regression coefficient (with the other coefficients for social factors being small). 
%
%% Table: regression coefficients
%\begin{table}[htbp!]
%\centering
%\includegraphics[width=\linewidth]{figs/regression_table_v2.pdf}
%\caption{{\bf Predicting  node degrees using  combinations of demographic, socioeconomic and infrastructure information.}
%%
%{\color{revisions}
%We report R-squared scores and regression coefficients for six different regression models that consider the prediction of either {\bf (a)} out-degrees versus social factors, {\bf (b)} in-degrees versus number of POIs, and {\bf (c)} in- and out-degrees versus all features. These models are fit to empirical data during the baseline weeks.
%%
%In panel {\bf (c)}, we also report how these R-squared scores and regression coefficients changed during the storm week.
%%
%%See text for further discussion.
%%comparing certain social factors and number of POIs to out-degree and in-degree. The models use high-movement categories and the baseline mean-degree. The bottom panel shows the change in values for models using the storm week (storm - baseline). Note, the data was not divided into training and testing groups, to better analyze relationships and compare to r-values in Section \ref{sec:mixing}. We see the best model for both in-degree and out-degree use all social factors and a breakdown of movement categories. 
%}
%}
%\label{tab:regression}
%\end{table}
%
Finally, in the last two columns of Figure \ref{fig:6}(d), we report how  $R^2$  and regression coefficients changed when the multivariate regression models are fit to data during the storm week. 
We first consider the models that predict out-degrees during the storm (change for model 5), finding that models'  regression coefficients significantly change (almost always decreasing in magnitude); however, 
 $R^2$ changed by less than 1\%. That is, outward movements can be predicted with nearly the same accuracy during the storm week.
We next consider the models that predict in-degrees during the storm (change for model 6) for which  $R^2$ decreased by 4\%, and we see a decrease in the importance of industry information. 
Lastly, while not included in the table, we also studied how the models 1-4 changed during the storm week. We found that during the storm,  $R^2$ dropped by less than 1\% when predicting out-degree vs social factors, but that they dropped by 12-13\% when predicting in-degree vs number of POIs. %For both situations the regression coefficients had a decrease similar to (c).

In conclusion, inward movements into \revrev{census tracts} are generally harder to predict than outward movements (e.g., $R^2$  are much smaller for in-degree versus out-degree) and their prediction is also much more impacted by the storm (e.g., the drop in  $R^2$   is much greater).
}

%------------------------------------------------------------------------------------------------------------------------%------------------------------------------------------------------------------------------------------------------------
\section{\rev{Discussion}}\label{sec:discussion}%------------------------------------------------------------------------------------------------------------------------%------------------------------------------------------------------------------------------------------------------------

% Summary
In this work, we studied human mobility using time-varying, multilayer networks in which edges encode observed movements between spatial regions (i.e., census tracts) and network layers encode different movement categories that were defined according to industry sector (e.g., visitations to schools,   hospitals, and grocery stores).  
While multilayer networks were  utilized to encode different modes of transportation (e.g., roadways versus metro lines) in \revrev{previous} human mobility research   \cite{de2014navigability,taylor2015topological,chodrow2016demand}, our study leveraged them to encode different industry sectors of movements and investigated human mobility changes in different layers during a major disaster\revrev{: Winter Storm Uri.}
By considering mobility patterns by industry sector, we  gained complementary insight about how the same storm can have different impacts on human movements in different industry sectors.

Focusing on Harris County, TX, we found  that people reduced their movements to ambulatory healthcare services, restaurants, and schools but prioritized movements to grocery stores and gas stations.
We additionally studied the predictability of inward and outward movements for \revrev{census tracts} using information about their demographic, socioeconomic, and infrastructure characteristics. We found that as compared to outward movements (i.e., out-degrees), inward movements (i.e., in-degrees) are harder to predict especially during the storm. These  insights into the reduction, prioritization, and predictability of human movements during Winter Storm Uri could be useful for supporting  the decisions of policy makers and emergency responders during extreme weather events. 

\revrev{
More broadly, this case study illustrates the effectiveness of our methodology for detecting and characterizing mobility shifts, suggesting that it will be useful  for diverse scenarios even beyond weather events. 
%even those lacking top-down directives from government officials.
%methodology presented herein has broad potential applications beyond weather events. O
%
It is also worth noting that the mobility changes observed herein reflect a combination of  influences including voluntary choices, power outages, and government edicts (e.g., building closures and travel recommendations). While we cannot decouple these effects for storm Uri, 
%it is interesting to consider whether it is possible, in principle, to disentangle the various drivers for human-movement change (whether for storms or other events). 
%remains a difficult open problem
%
our techniques should be useful for diverse scenarios including those with or without top-down directives from government officials.
%Nevertheless, this case study demonstrates the effectiveness of our proposed methodology, suggesting its effectiveness for detecting and characterizing mobility shifts for diverse scenarios even those lacking top-down directives from government officials. Moreover, 
Moreover, it is interesting to consider whether it is even possible, in principle, to disentangle these various drivers for human-movement change. If so, stratifying mobility networks by industry type (as we have proposed) could lead to  fruitful directions to address this open challenge.
}

%In this regard, our approach could be used to detect shifts in movements based purely on individual's choice, when there is no official recommendations or closures.}
%\revrev{We note that the decrease in human mobility can be due to voluntary choice or government edicts (including building closures and travel recommendations), and we can not decouple these effects in our study. In this regard, our approach could be used to detect shifts in movements based purely on individual's choice, when there is no official recommendations or closures.}

%Winter Storm Uri dramatically affected movement patterns for Harris County, TX. We studied the impact on different movement categories, according to an industry classification hierarchy, utilizing human mobility data from anonymized mobile device locations to create a time series of stratified networks for the movement categories, comparing the week of the storm to a set of baseline weeks. We examined the correlation between movement direction and factors including number of POIs and six socioeconomic and demographic attributes using statistical tests. Our results offer information about the impact of the storm on movement categories and population groups, compared to a baseline. This study provides an industrial and residential perspective of the various impacts of the 2021 winter storm. 

%We conclude by highlighting that recent years have given rise to numerous techniques

%in this paper, we focused on studying the number of edges and the node degrees for individual (i.e., uncoupled) network layers. These are arguably 

We end by highlighting a few \revrev{additional} future  directions \revrev{for research}.  In this \revrev{paper}, we constructed network models in which weighted edges encode the numbers of observed movements, and it would be interesting to study other types of networks including those where edge weights are normalized (e.g., by the density of devices) or reflect geospatial information (e.g., distances between \revrev{census tracts}, \revrev{census tract} land areas, and spatial partitioning biases).
\rev{We also studied a set of social factors that did not include age-related information (i.e., which were removed during  \revrev{variance inflation factor tests to remove correlated drivers and} ensure statistical rigor as discussed in Sec.~\ref{sec:regression}), and it would be interesting to investigate how different age groups were impacted in future research. However, it is also known that mobile-device data under represent older populations \cite{li2024understanding, chang2022role}, posing a major challenge to such a study.}
%
%did not study the impact on age-related social factors. This is due to the data being aggregated to the \revrev{census tract} level and not providing information on individual mobile users. It would be interesting to study the impact on 65 and older populations, but the dataset doesn't provide the necessary level of detail and mobile-device data often under represents older populations \cite{li2024understanding, chang2022role}. }
%
One could also broaden the characterization of the anomalous week by using quantification methods besides z-scores and by considering different choices for selecting baselines that incorporate, e.g., seasonal and annual trends.
\rev{
Incorporating data from multiple years is another direction; the 2020 data is significantly impacted by COVID-19, but one could explore techniques to debias its impacts.}
Also, it would be interesting to compare our retrospective study of human mobility during Winter Storm Uri to  movements observed in other extreme weather events.
\rev{Finally, further study of the predictability of specific industry categories would benefit the understanding of the increased demand on certain industries as well as provide insight into the varied predictability of different  categories.}
%

%Additional information about deaths and accidents could also provide context to explore certain NAICS categories and social factors to provide better assistance during weather crisis. 
%

% Note that age-related features were removed due to correlation with population. We believe this is due to the fact that our dataset is based on census tracts and does not contain specific age details for individual users. It is also known that older populations are often underrepresented in mobile-device based data \cite{li2024understanding, chang2022role}.}
%

%---------------------------------------------------------------------------------
%---------------------------------------------------------------------------------
\section{Methods}\label{sec:methods}
%---------------------------------------------------------------------------------
%---------------------------------------------------------------------------------

\rev{\subsection{Construction of multilayer networks}}\label{sec:construct}

%\rev{
%We first introduce our approach to modeling human movement between \revrev{census tracts}, focusing on Harris County, TX.
%begin by introducing how the networks are constructed from our dataset.
%}
We study a dataset of weekly human movements from the data provider SafeGraph. 
%, that describes visits from home locations to destination locations. 
The approximately 700 GB dataset is collected based on the GPS locations of  opt-in smart mobile devices (mostly smartphones), and captures weekly movements of people from a home location, recorded at the corresponding census block group, to a destination location, i.e., a specific point of interest (POI).  We use the Python library \textbf{safegraph\_py} to process and prepare the data into this graph-structured format and computations were implemented on the NCAR-Wyoming Supercomputing Center.

%Observe that t
The original SafeGraph data can be encoded by a time-varying, bipartite graph $G(t)$, for $t=1,\dots, T$, where $T$ is the total number of weeks studied. 
Each graph $G(t)$ is composed of weighted, directed edges that encode the number of observed movements between a source \revrev{census block group} and destination POI.
We denote $\mathcal{H} = \{CBG_i\}_{i=1}^H$ to be a set of source nodes (i.e., the home \revrev{census block groups} associated with mobile devices), where $H$ is the total number of \revrev{census block groups} in the studied area, and let $\mathcal{P} = \{POI_i\}_{i=1}^P$ be a set of destination nodes (i.e.,  the set of POIs), where $P$ is the total number of POIs. 
POIs are identified using SafeGraph's Placekey system, a universal location identifier that combines a geospatial encoding system with a unique POI identifier that provides information including the name, latitude and longitude, business details, and industry classification.  
For privacy reasons, SafeGraph omits sparse data in which fewer than four visitors are recorded in a given week from any home \revrev{census block group} to a POI. 
Each edge $(i,j)$ in $G(t)$ has a weight $B_{ij}(t)$ that encodes the number of observed movements from $CBG_i$ to $POI_j$ during week $t$. Equivalently, each graph $G(t)$ can be encoded by a weighted adjacency matrix, $B(t) \in \mathbb{R}^{H \times P}$.
%
%Note that \revrev{census block groups} are a neighbor-hood level spatial unit and several \revrev{census block groups} comprise a more-commonly known Census Tract (\revrev{census tract}). %We will use \revrev{census tracts} for our spatial discretization for c. For each \revrev{census tract} and \revrev{census block group} we compute the latitude and longitude of the centroid. 
%
%
% explain networks

\rev{We first discuss the coarse-graining of SafeGraph data to the spatial resolution of census tracts. }   We define $\mathcal{C} = \{CT_i\}_{i=1}^C$ to be a set of \revrev{census tracts} of interest, where $C$ is the total number of \revrev{census tracts} ($C=786$ for Harris County) and let  $\mathcal{H}_i$ and $\mathcal{P}_i$ denote, respectively, the \revrev{census block groups} and POIs within $CT_i$. Then the combined observed movement from \revrev{census block groups} in $CT_i$ to POIs in $CT_j$ during week $t$ is given by 
\begin{equation}\label{eq:Aij}
A_{ij}(t) = \sum_{i' \in \mathcal{H}_i, j' \in \mathcal{P}_j} B_{i'j'}(t).
\end{equation}
The remainder of this study examines time-varying networks encoded by square adjacency matrices $A(t)$ that are  size $C\times C$.
Finally, for each network, we define $d^{out}_i(t) = \sum_{j} A_{ij}(t)$ to be the out-degree---that is,  a measure for all movements during week $t$  that leave   $CT_i$---and $d^{in}_i(t) = \sum_{j} A_{ji}(t)$ to be the in-degree---that is, a measure for all movements to POIs within $CT_j$ during week $t$.
For the mobility network in Figure \ref{fig:1}(b), for example,  in- and out-degrees for $CT_1$   would be  $d^{in}_1(t) = A_{11}(t) $ and $d^{out}_1 (t) = A_{11}(t) + A_{12}(t) + A_{13}(t)$.

%Our study integrates a combination of human movement data and socioeconomic indicators to study mobility patterns in Harris County, TX and how they are impacted by weather anomalies. 
%
%We use SafeGraph \cite{safegraph} mobility data, which tracks movements based on GPS locations, and integrate this with demographic and socioeconomic data from the U.S. Census Bureau \cite{census}. The analysis involves creating networks that represent movement patterns between Census Tracts and Points of Interest, stratifying these networks by movement type and time. We quantify the impact of external events, such as a storm, on mobility using z-scores that measure event rareness. Additionally, we explore correlations between movement patterns and key socioeconomic factors using regression analysis. Combined, this methodological framework allows for a comprehensive understanding of the relationships between mobility, the impacts of winter storms, and social characteristics in the region.

\rev{We next discuss behavioral-stratification of the networks into layers encoding movements to industry categories based on the NAICS heirarchy.}
 Letting $\mathcal{N}$ denote the set of NAICS codes with a fixed number of digits, for each   $n \in \mathcal{N}$ we  define $\mathcal{P}^{(n)}_j$ as the set of POIs within $CT_j$ having NAICS code $n$. The  network layers' adjacency matrices are  then obtained by
%
%\begin{equation}\label{eq:Anij}
$A^{(n)}_{ij}(t) = \sum_{i' \in \mathcal{H}_i, j' \in \mathcal{P}^{(n)}_j} B_{i'j'}(t).$
%\end{equation}
%
%In this work, we often focus on the subset of movement categories associated with the most movements.
For each network layer $n$ and $CT_i$, we define the time-varying in- and out-degrees by  $d^{in,(n)}_i(t) = \sum_{j} A^{(n)}_{ji}(t)$ and  $d^{out,(n)}_i(t) = \sum_{j} A^{(n)}_{ij}(t)$, respectively.
Finally, note that summing $A^{(n)}_{ij}(t)$ over all possible $n\in\mathcal{N}$ recovers the adjacency matrix  $\sum_n A^{(n)}(t) = A(t)$ of the original, non-stratified network (i.e., also   called the layer-aggregated network). 
One can similarly obtain the in- and out-degrees for the network encoding all movements by summing over the network layers: $d^{in}_i(t) = \sum_{n} d^{in,(n)}_i(t) $ and $d^{out}_i(t) = \sum_{n} d^{out,(n)}_i(t) $.

\rev{\subsection{Determining threshold for high-movement categories}}\label{sec:threshold}
%===================================================

%We can use movement categories to better study human movements, by looking into the infrastructure vs the number of movements. 

For many of the NAICS categories, the numbers of observed movements can be much smaller than those for other categories. Therefore, throughout this paper we will often focus on the categories with the most movements. 
In Figure \ref{fig:2}(c), we show the number of POIs by NAICS category for Harris County (left), 
%the widths of the wedges represent the fraction of POIs for each category for all of Harris County, 
$\sum_{j} \left| \mathcal{P}^{(n)}_j \right|$, whereas in Figure \ref{fig:2}(c) we depict the \revrev{total}  movements for each category (right). That is, for each NAICS code $n\in\mathcal{N}$, we compute $m^{(n)}(t) = \sum_{i,j} A_{ij}^{(n)}(t)$ to be the total movement during week $t$ and $M^{(n)}= \sum_{t}m^{(n)}(t)$ to be the total movement across the study duration. \revrev{The coloration in each bar depicts the finer subcategories (i.e., NAICS codes with 3-4 digits).}
% $M^{(n)}= \sum_{t}m^{(n)}(t)$, where
%\begin{equation}\label{eq:movement}
%m^{(n)}(t) = \sum_{i,j} A_{ij}^{(n)}(t),
%\end{equation}
%is the total movements for category, $n$, during week, $t$. 
%
We then divide  the 2-digit NAICS categories into a set $\mathcal{N}_{high} = \left\{ n \mid M^{(n)} \geq 10^{-6} \right\} = \left\{44, 45, 53, 61, 62, 71, 72 \right\}$ of high-movement categories and a set $\mathcal{N}_{low} = \left\{ n \mid M^{(n)} < 10^{-6} \right\}$ of low-movement categories.
\rev{The choice of $10^{-6}$ was selected as a natural partition of the dataset, since the total movements are much larger for educational services versus transportation services (i.e., $M^{(61)} = 1,462,340$ versus $M^{(48)} = 475,732$), which is the largest percentage change.}
\rev{The $M^{(n)}$ values of all 2-digits NAICS categories and a}  detailed breakdown of the categories into subcategories is shown in Supplementary Table 1. This provides overview of the hierarchy of layers and the total amount of movement in each layer for Harris County. The set $\mathcal{N}_{low}$ contains 16 categories and are either combined into a ``low-movement category''  or omitted from our study.

%===================================================
\subsection{Z-scores quantify movement-change significance}\label{sec:zscore}
%===================================================

% Introduction
%We quantify the storm's impact on total movements for different network layers by calculating z-scores.
% and examining whether the stchanges increase or decrease movements for a particular behavioral category. 
We quantify the storm's impact on total movements for different network layers by comparing the total movements $m^{(n)}(t)$ during the storm week to a baseline of weekly movement, and we use z-scores to measure deviation from typical behavior. We apply this approach to different network layers to identify which movement categories undergo statistically significant change.
 %that occur for different . 
%
%To further explore movement patterns we also look into the direction of the movements, using node degree. We define out-degree as the movements from a home \revrev{census tract} and in-degree as the movements into a destination \revrev{census tract}, capturing movement direction between locations. This methodology provides a structured way to evaluate the influence of extreme events on human movement.
%
% z-score and baseline definition
%To assess the storm's impact on movement patterns, we compare total movements during the storm week to movements during other weeks and quantify the amount of disruption by calculating the z-score of the storm week relative to a baseline. 
After visually inspecting time series encoding weekly movements for several years of data (see Supplementary Figure 2), we select the baseline that consists of the six weeks prior to the storm,  $\mathcal{T}_{base} = \{2, 3, \dots, 7\}$. 
Letting $\mu$ and $\sigma$ denote the mean and standard deviation of $m(t)$ across $t\in\mathcal{T}_{base}$, we compute the z-score $Z(t) = \frac{m(t) - \mu}{\sigma}$ to study the storm's impact on all movement categories.
% total movements deviate from the baseline mean
%\frac{M_{base}}{| \mathcal{T}_{base} |}$, where $M_{base} = \sum_{t \in \mathcal{T}_{base}} m(t)$
Similarly, for each movement category $n$ we let
 $\mu^{(n)}$ and $\sigma^{(n)}$ denote the mean and standard deviation of $m^{(n)}(t)$ across $t\in\mathcal{T}_{base}$ and compute the z-score 
 \begin{equation}\label{eq:zscore}
Z^{(n)}(t) = \frac{m^{(n)}(t) - \mu^{(n)}}{\sigma^{(n)}}. 
\end{equation}
We compare z-scores across movement categories in Section \ref{sec:impact}.

\subsection{Regression analysis relates movements to infrastructure, demographic, and socioeconomic information}\label{sec:regression}
%===================================================
In Section~\ref{sec:census}, we study the in- and out-degrees for network layers and investigate how these structural properties relate to infrastructure information (i.e., derived from the POIs) as well social factors including demographic and socioeconomic information.
% Explain census data and quartiles
To this end, we gathered  data from the U.S. Census Bureau, accessing tables from the 2010 American Community Survey and filtering for year 2019 and   \revrev{census tracts} in Harris County, TX. For each \revrev{census tract}, we assembled data for 13 social factors: population (B01003), population density, under 18 (DP05), under 5 (DP05), income (B19013), unemployment (DP03), poverty rate (S0601), non-white percentage (B02001), non-hispanic and non-black percentage (S0601),  owner occupied percentage (B25003), renter occupied percentage (B25003), education level (S0601).

To identify which social factors have the strongest correlation with the networks' in- and out-degrees, we conduct a multivariate linear regression.
% that predicts in and out-degrees within each layer (or across layers) based on social factors.
Noting that some social factors are correlated, provide redundant information, and cause regression instability,
we sought to obtain a smaller set of social factors. Specifically, we conducted variance inflation factor (VIF) \rev{tests} to identify and remove variables having multicollinearity, which can distort regression coefficients and reduce the model's reliability. For each social factor, $X^{(i)} \in \mathbb{R}^{786}, i=1,\dots,13$, we calculated $VIF_i = 1/(1-R^2_i)$,  where
\begin{equation}\label{eq:r2}
R_i^2 =  1 - \frac{\text{residual sum of squares}}{\text{total sum of squares}} = 1 - \frac{\sum_{j=1}^{786} \left(X^{(i)}_j - \hat{X}^{(i)}_j\right)^2}{\sum_{j=1}^{786} \left(X^{(i)}_j - \bar{X}^{(i)}\right)^2},
\end{equation}
$R^2_i$ is called the coefficient of determination.
%A high $R^2_i$ means $X^{(i)}$ is strongly correlated with the other predictors, leading to a higher VIF. 
%for the regression equation of each social factor against the others, given by
Here, each $\hat{X}^{(i)}$ is a predicted value from regression and $\bar{X}^{(i)}$ is the mean across observed values. 

\rev{In Table \ref{tab:vif}, we summarize our results for several VIF tests that were were run using the \textit{statsmodels} module in Python. For Test 1, we used 12 predictors that span three categories: demographic, socioeconomic, and race/ethnicity.  In Test 2, we removed a predictor from each category, which were selected based on having a high VIF value in Test 1.
%as well as intuitive correlations (e.g. under 18 population and school enrollment). 
For Test 3, we removed two predictors with the highest VIF values from Test 2.  Finally, Test 4 showed that the VIF values of all remaining predictors were less than 5 (which is a standard threshold in VIF tests). The final six predictors are: population, population density, income, unemployment \%, poverty rate \%, and non-white \%. 
}
%\todo{Do we want Table 1 closer to this paragraph? }
% Note that age-related features were removed due to correlation with population. We believe this is due to the fact that our dataset is based on census tracts and does not contain specific age details for individual users. It is also known that older populations are often underrepresented in mobile-device based data \cite{li2024understanding, chang2022role}.}
%

\begin{table}[htbp!]
\centering
\begin{tabular}{l l c c c c}
\hline
\textbf{Category} & \textbf{Predictor} & \textbf{Test 1} & \textbf{Test 2} & \textbf{Test 3} & \textbf{Test 4} \\
\hline
\multirow{5}{*}{Demographic} 
 & Population & 5.318 & 4.785 & 3.531 & 3.375 \\
 & Population Density & 3.308 & 3.292 & 2.687 & 2.672 \\
 & 65 years and over (\%) & 8.458 & 5.516 & 5.444 & -- \\
 & Under 18 (\%) & 84.755 & -- & -- & -- \\ 
\hline
\multirow{5}{*}{Socioeconomic} 
 & Income & 14.488 & 6.156 & 5.177 & 2.675 \\
 & School Enrollment (\%) & 63.847 & 12.666 & -- & -- \\
 & Unemployment (\%) & 4.884 & 4.814 & 4.535 & 4.255 \\
 & Poverty Rate (\%) & 8.111 & 7.557 & 4.021 & 3.933 \\
 & Renter Occupied (\%) & 18.919 & 8.367 & -- & -- \\
 & Owner Occupied (\%) & 28.744 & -- & -- & -- \\
\hline
\multirow{2}{*}{Race/Ethnicity} 
 & Non-White (\%) & 6.201 & 5.609 & 5.038 & 4.849 \\
 & Non-Hispanic and Non-Black (\%) & 10.992 & -- & -- & -- \\
\hline
\end{tabular}
\caption{\rev{{\bf \revrev{Variance inflation factor (VIF)  tests} to identify and remove variables having multicollinearity.} For Test 1, we used 12 predictors from 3 categories and resulted in most VIF values being greater than 5 (which is a standard threshold). Then we conducted successive tests to remove the highest VIF score from each category (while leaving at least one per category).
%For Test 2, we removed 1 predictor in each category based on having high VIF values and intuitive correlations (e.g. school enrollment and under 18). For Test 3, we removed the predictors with the highest VIF values from the demographics and economic categories. 
For the final test (Test 4), each predictor in Test 3 was removed one at a time to identify a set of factors such that no VIF values were greater than 5.
}}
\label{tab:vif}
\end{table}

% We successively removed variables 
% %with the  highest VIF 
% {\color{revisions} according to VIF values and similar categories, e.g. population predictors,}
% (implemented using the \textit{statsmodels} module in Python) until all VIFs were less than five, leaving the following six social factors and VIFs shown in Table \ref{tab:vif}.
%The \textit{statsmodels} module in Python was used to perform the VIF tests. We then used these six variables in a multivariate linear regression to create a predictive model for out-degrees for the baseline weeks. 

%Table: VIF values
% \begin{table}[ht]
% \centering
% \begin{tabular}{c|l|l}
% $X^{(i)}$ & Description & $VIF_i$\\
% \hline
% $X^{(1)}$ & Population & 3.38 \\
% $X^{(2)}$ & Population Density & 2.67 \\
% $X^{(3)}$ & Income & 2.67 \\
% $X^{(4)}$ & Non-White Percentage & 4.85 \\
% $X^{(5)}$ & Poverty Rate & 3.93 \\
% $X^{(6)}$ & Unemployment Rate & 4.25 \\
% \end{tabular}
% \caption{Variance inflation factors (VIFs) for 6 selected socioeconomic and demographic variables.}
% \label{tab:vif}
% \end{table}
%

% Describe multillinear regression and VIF
In addition to the social features above, we also include infrastructure features for each \revrev{census tract} in the form of number of POIs. We look at total number of POIs for each \revrev{census tract} as well as a breakdown of number of POIs for the high-movement categories, $\mathcal{N}_{high}$, shown in Figure \ref{fig:2} (c). We highlight that we never include both the total number of POIs and the stratified POIs into categories, since that would  introduce collinearity into the model (i.e.,   the total of POIs equals the summation over POIs in different categories).
Each model defines a relationship between either in- or out-degrees and a set of features, $\left\{ X^{(i)} \mid i \in S \right\} $, where $S$ is a set of select indexed features (possibly including demographic, socioeconomic,  and infrastructure information). 
\rev{The features were standardized using z-score normalization (mean = 0, standard deviation = 1) prior to regression to ensure comparability of coefficients across features.} 
We then fit linear regression models of the form
\begin{equation}\label{eq:regression}
Y_S = \epsilon + \beta_0 + \sum_{i \in S} \beta_iX^{(i)}
\end{equation}
using the \textit{scikit-learn} module in Python and test their fitness by examining  associated R-squared scores, similar to Equation \ref{eq:r2}. 
%Looking at various subsets of features allows us to compare models and   relationships between movement patterns.

%------------------------------------------------------------------------------------------------------------------------%------------------------------------------------------------------------------------------------------------------------
% OTHER STUFF
%------------------------------------------------------------------------------------------------------------------------%------------------------------------------------------------------------------------------------------------------------

%\noindent LaTeX formats citations and references automatically using the bibliography records in your .bib file, which you can edit via the project menu. Use the cite command for an inline citation, e.g.  \cite{Hao:gidmaps:2014}.
%
%For data citations of datasets uploaded to e.g. \emph{figshare}, please use the \verb|howpublished| option in the bib entry to specify the platform and the link, as in the \verb|Hao:gidmaps:2014| example in the sample bibliography file.
\section*{Acknowledgements}

This work is supported by the U.S. National Science Foundation under Grant No. DMS-2401276.
%, BCS-2117771 and BCS-2318206. 
Any opinions, findings, and conclusions or recommendations expressed in this material are those of the authors and do not necessarily reflect the views of the
National Science Foundation. The authors also thank SafeGraph for providing anonymized mobile
phone location data and the Jay Kemmer WORTH Institute for seed funding.
%
%\section{Author contributions statement}
%
%Must include all authors, identified by initials, for example:
%A.A. conceived the experiment(s),  A.A. and B.A. conducted the experiment(s), C.A. and D.A. analysed the results.  All authors reviewed the manuscript. 

\section*{Additional Information}
%To include, in this order: \textbf{Accession codes} (where applicable); \textbf{Competing interests} (mandatory statement). 
\textbf{Author contributions:} All authors developed the study and contributed regularly. MB, AK and FA implemented the analyses. MB led the manuscript writing, which was read and approved by all authors.
\\
\noindent
\textbf{Code availability:}
Code base and processed network data can be found at
\\
 \href{https://github.com/NSF-ATD-MobilityNetwork/human_mobility}{https://github.com/NSF-ATD-MobilityNetwork/human\_mobility}. 
 \\
\noindent
\textbf{Data availability:}
Original SafeGraph data is proprietary and researchers can contact SafeGraph (www.safegraph.com). NAICS classifications are made available by the US Census (www.census.gov/naics).
%The data that support the findings of this study are available from SafeGraph but restrictions apply to the availability of these data, which were used under license for the current study, and so are not publicly available. Data are however available from the authors upon reasonable request and with permission of SafeGraph (\url{https://www.safegraph.com/}).
\\
\noindent
\textbf{Competing Interest Statement:}
The authors declare that they have no competing financial interests or personal relationships that could have appeared to influence the work reported in this paper.

%The corresponding author is responsible for submitting a \href{http://www.nature.com/srep/policies/index.html#competing}{competing interests statement} on behalf of all authors of the paper. This statement must be included in the submitted article file.

%\begin{figure}[ht]
%\centering
%\includegraphics[width=\linewidth]{stream}
%\caption{Legend (350 words max). Example legend text.}
%\label{fig:stream}
%\end{figure}
%
%\begin{table}[ht]
%\centering
%\begin{tabular}{|l|l|l|}
%\hline
%Condition & n & p \\
%\hline
%A & 5 & 0.1 \\
%\hline
%B & 10 & 0.01 \\
%\hline
%\end{tabular}
%\caption{\label{tab:example}Legend (350 words max). Example legend text.}
%\end{table}
%
%Figures and tables can be referenced in LaTeX using the ref command, e.g. Figure \ref{fig:stream} and Table \ref{tab:example}.

\bibliography{refs.bib}

\clearpage



\section*{Supplementary material}

# 5 Supplementary Material: Appendices

## 5.1 Detailed breakdown of POI and movement categories into sub-categories

In Section 2.2, we discussed the stratification of movements to POIs into industry sectors, and the breakdown of POIs and movements into these sectors was illustrated in Figure 3. In Figure 9, we present additional detail about the breakdown, with the stratification of POIs and movements shown in the left and right columns, respectively. Line widths indicate either the number of POIs (left) or movements (right). To simplify the visualization, categories with small total movement counts are combined into a 'Low Movement' category. Additionally, categories with zero counts are omitted from the figure to maintain clarity and avoid clutter.

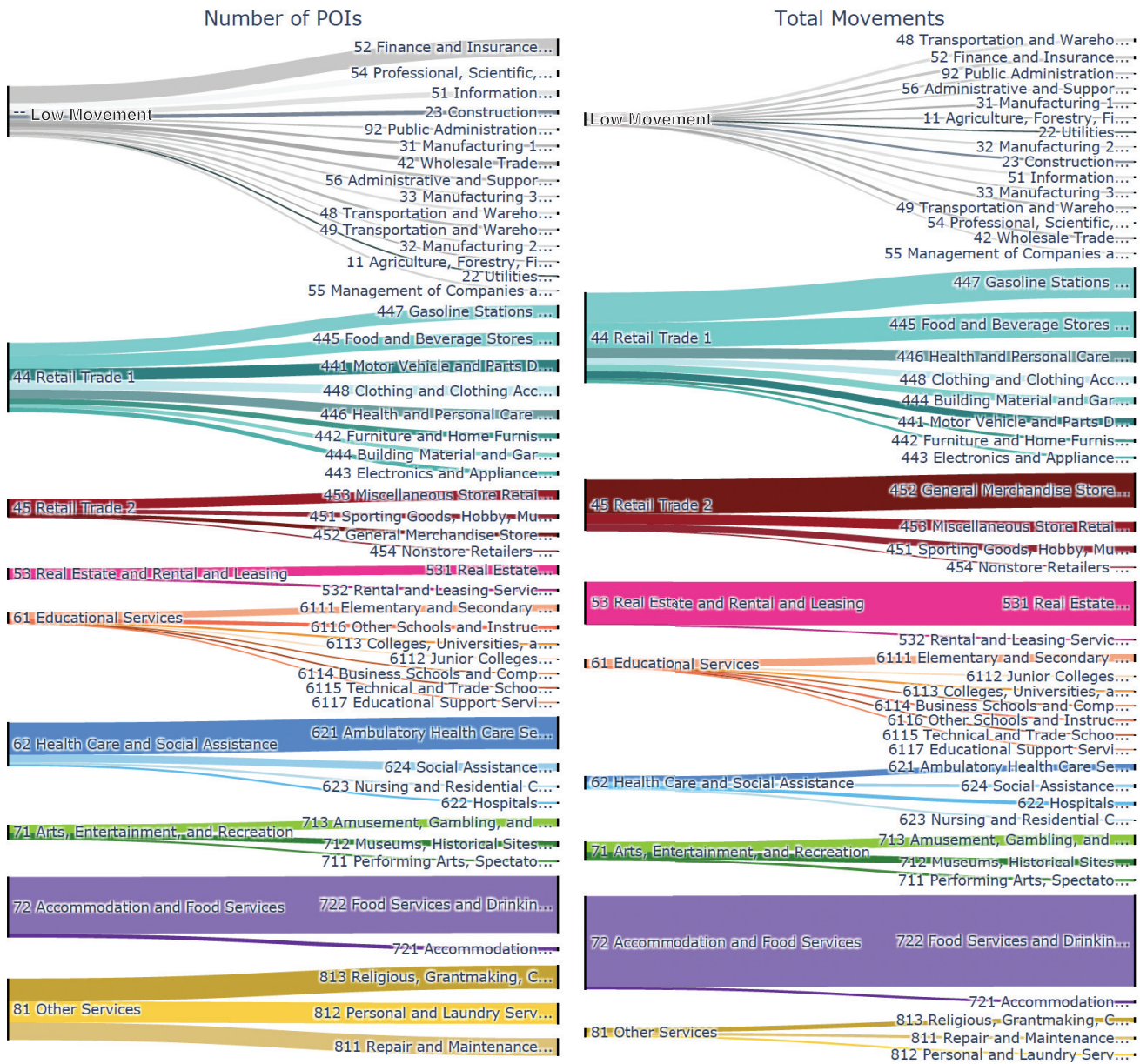

**Figure 9.** Detailed summary for the breakdown of POIs and observed movements categories and sub-categories. Extending Figure 2, we provide a detailed description for the breakdown of movement categories into subcategories. The line widths are proportional to the number of number of POIs (left) and the total movements (right) in each industry category and subcategory. Categories with few total movements are combined into a group call 'Low Movement'.

## 5.2 Multi-year, total-movement time series for select categories

In Section 2.3, we discuss our use of z-scores to quantify anomalous levels of total movement. Our approach involved comparing total movements during the storm week to a baseline distribution, which was informed by examining time series for total movements spanning the years 2018-2021. We provide such time series data in Figure 10, where one can observe in years 2018, 2019, and 2021 that there is a general trend of weekly increases in total movement as the winter turns to spring, but this does not occur in 2020 due to the COVID 19 pandemic. The red shading marks the same calendar week in each year as the 2021 storm. These complex seasonal patterns and the unique event in 2020 influenced our decision to select the baseline as the six weeks prior to the storm. This selection as further supported by examining time series for 2021 alone, as shown in Figure 11.

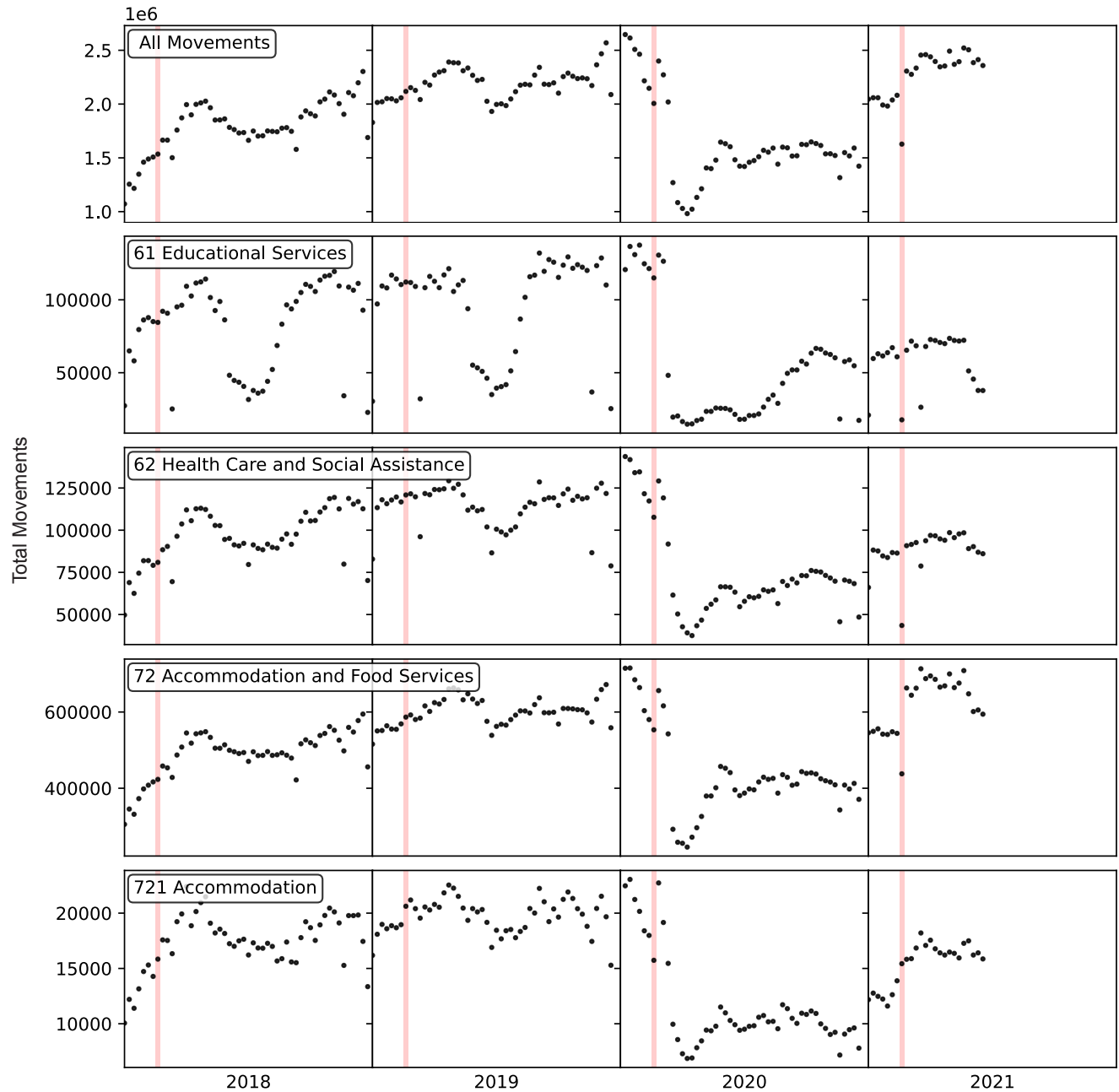

**Figure 10. Multi-year 2018–2021 time series for total movements.** The top row shows total movements across all movement categories, whereas the other rows depict time series for the categories that were most impacted by the storm. The bottom row shows movements for accommodation (721) to observe the seasonal increase every year around the same time. Observe all time series significantly drop in 2020 due to the COVID19 pandemic.

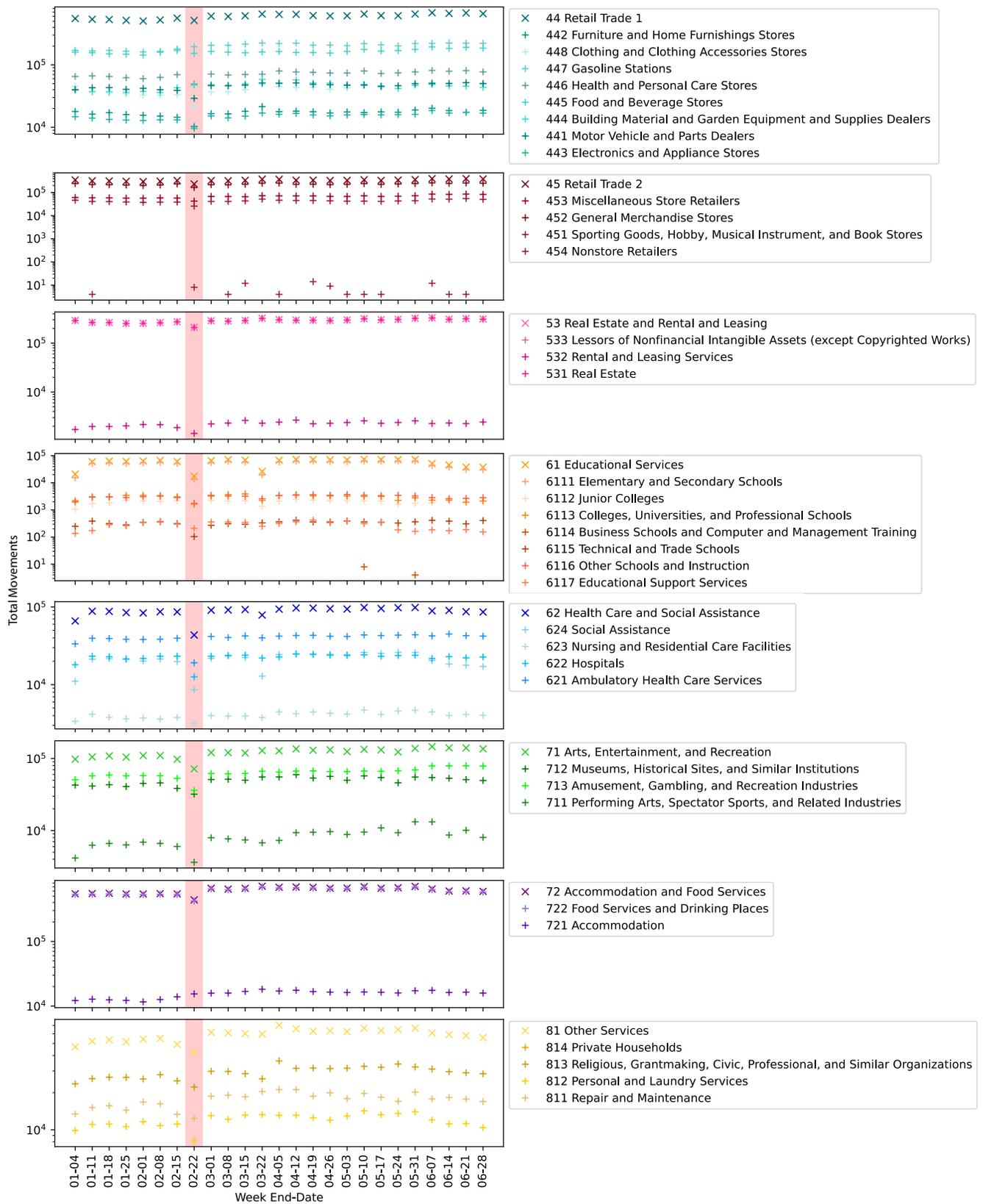

**Figure 11. Total movements for high-movement NAICS categories and subcategories in 2021.** The storm week is highlighted by the red shading.

## 5.3 Degree distributions for different movement categories

In Section 3.2, we studied the storm's impact on CTs' inward and outward flows (i.e., nodes' in- and out-degrees), and in Figure 7(a)–(b) we plotted the degree distributions for the network encoding all movement categories. Extending this study, in Figure 12 we depict the degree distributions for the network layers encoding high-movement behavioral categories. Separate panels depict these distributions during the baseline weeks and the storm week. Most distributions approximately follow a straight line in a log-log scale, and the degrees generally shift toward smaller values during the storm.

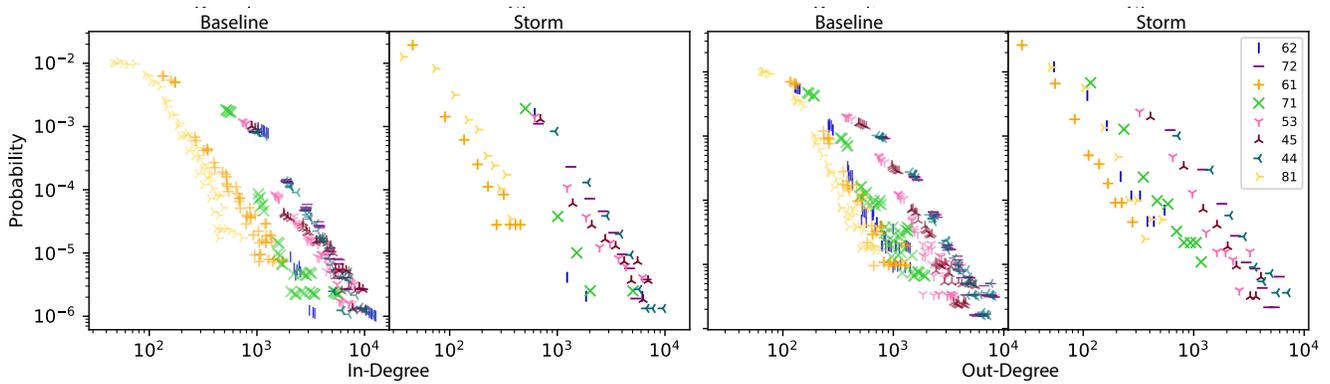

**Figure 12. Degree Distributions for Network Layers.** Focusing on network layers associated with high-movement categories, we plot the distributions of in-degrees (left) and out-degrees (right) for both the baseline weeks and the storm week.



\end{document}